\documentclass[10pt,twocolumn,twoside]{IEEEtran}

\ifCLASSINFOpdf
   \usepackage[pdftex]{graphicx}
   \graphicspath{{./figs/}}
\else
\fi

\usepackage{amsmath,amssymb,amsthm}
\usepackage{algorithm,algorithmic}

\usepackage{array}

\ifCLASSOPTIONcompsoc
  \usepackage[caption=false,font=normalsize,labelfont=sf,textfont=sf]{subfig}
\else
  \usepackage[caption=false,font=footnotesize]{subfig}
\fi
\usepackage{url}

\usepackage{float,stackrel,adjustbox,tabularx,multirow,tikz,nicefrac,diagbox,hhline,color,colortbl,xcolor,psfrag,pstool}

\usetikzlibrary{spy,arrows.meta,arrows,calc}

\usepackage[utf8]{inputenc}
\usepackage{fancyhdr,amsmath,amsfonts,amsthm,amssymb,graphicx,tikz,nicefrac,algorithmic,enumerate,ulem}
\usepackage{cite,lastpage,xcolor,verbatim,color}

\DeclareMathOperator{\E}{\mathbb{E}}
\DeclareMathOperator{\bbP}{\mathbb{P}}

\newcommand{\T}{^{\mbox{\tiny \sf T}}}

\newcommand{\R}{\mathbb{R}}
\newcommand{\G}{\mathcal{G}}
\newcommand{\V}{\mathcal{V}}
\newcommand{\Ed}{\mathcal{E}}
\newcommand{\J}{\Lambda}
\newcommand{\N}{\mathcal{N}}
\newcommand{\bbN}{\mathbb{N}}
\newcommand{\xad}{\hat{x}^{\rm ad}}
\newcommand{\ead}{e^{\rm ad}}

\newcommand{\sad}{\Sigma^{\rm ad}}
\newcommand{\tr}{\mathrm{tr}}
\newcommand{\aep}[1]{\textit{asymptotically} $#1$-\textit{protected}}
\newcommand{\ep}[1]{$#1$-\textit{protected}}

\newcommand{\diag}[1]{\mathrm{diag}[#1]}

\renewcommand{\baselinestretch}{0.98}

\theoremstyle{definition}

\newtheorem{definition}{Definition}
\newtheorem{rem}{Remark}
\newtheorem{corr}{Corollary}

\newtheorem{pro}{Proposition}

\newtheorem{theorem}{Theorem}
\newtheorem{lemma}{Lemma}

\title{Multi-Agent Consensus Subject to Communication and  Privacy Constraints}
\author{
	Dipankar Maity and
	Panagiotis Tsiotras
	\thanks{D. Maity is an Assistant Professor in the Electrical and Computer Engineering department at the University of North Carolina Charlotte,  NC, 28223, USA. Email:
		{\small dmaity@uncc.edu}}%
	\thanks{P. Tsiotras is the David and Andrew Lewis Chair Professor with the Guggenheim School of Aerospace Engineering and the Institute for Robotics and Intelligent Machines, Georgia Institute of Technology, Atlanta,
		GA, 30332-0150, USA. Email:
		{\small tsiotras@gatech.edu}}
	}

\begin{document}


\maketitle


\begin{abstract}
    We consider a multi-agent consensus problem in the presence of adversarial agents. 
    The adversaries are able to listen to the inter-agent communications and try to estimate the state of the agents. 
    The agents have a limited bit-rate for communication and are required to quantize the transmitted signal in order to meet the bit-rate constraint of the communication channel.
    We propose a consensus protocol that is \textit{protected} against the adversaries, i.e., the expected mean-square error of the adversary state estimate is lower bounded.
    In order to deal with the bit-rate constraint, we propose a dynamic quantization scheme that guarantees \textit{protected consensus}.
\end{abstract}


\section{Introduction}
Consensus protocols, designed to ensure that all agents in a network come to an agreement, have been widely used in multi-robot systems \cite{alonso2016distributed, ren2008distributed},  satellite  altitude  alignment \cite{dimarogonas2009leader,wang1996coordination}, automated highway systems \cite{fagiolini2008consensus,olfati2002distributed}, and several 
civilian and military applications,  including target tracking and source localization \cite{kokiopoulou2010distributed,olfati2005consensus}. 
Beyond multi-agent systems in robotics, these problems also have a long history in the field of distributed computation \cite{lynch1996distributed,xu1996load}.

A typical consensus protocol \cite{olfati2004consensus}  consists of two components that are executed repeatedly. 
The first component defines an inter-agent communication scheme for each agent to send/receive data (typically the state of that agent)  to/from other agents in the network. 
The second component defines a computation scheme executed locally by each agent upon receipt of the data to update its own state.
These two processes are repeated until the states of all the agents become the same. 
In order for the agents to reach consensus, each agent follows the designed protocol including perfect transmission of its state. 
In realistic scenarios, perfect transmission of the data is difficult to achieve due to the limited data-rate (bit-rate) of the  communication channel. 
As the number of agents in the network grows, the available bit-rate per agent decreases. 
Therefore, the bit-rate constraint should  be incorporated explicitly when designing and analysing  consensus protocols for large-scale systems with dense communication architecture.

Privacy is important when the agents of the network deal with sensitive data. 
Recent technological advancements have made it possible to eavesdrop inter-agent communications and estimate the state of the network. 
Such an eavesdropper may not be a part of the network, and may remain undetected, while perfectly acquiring sensitive information regarding the agents. 
This provides an incentive for the agents to only consider consensus mechanisms that do not require them to share their true state values with the rest of the network.

\subsection{Previous Work}

To deal with limited bit-rate, quantized consensus protocols have been proposed. 
In \cite{li2013unified} the authors studied a consensus problem  with  quantization  and  time-delay and they showed that the states of the agents asymptotically converge to a \textit{practical consensus set}\footnote{
In practical consensus the states do not converge to a common value, rather they all remain within a small bounded domain.
}. 
In \cite{zhang2015periodic} the authors considered both quantization and event-triggered control to reduce the use of communication resources. 
The work of \cite{frasca2008average} showed that  randomized gossip algorithms can solve the \textit{average consensus}\footnote{
In average consensus each agent converges to a value that is the empirical average of all the agents' initial values. 
} problem with  quantized  communication.  
A  distributed algorithm in which the agents utilize probabilistically quantized information, i.e., dithered quantization, to communicate with each other is presented in \cite{aysal2008distributed}.
 A dynamic quantization scheme is presented in \cite{thanou2012distributed} that exploits the increasing correlation between the exchanged values by the agents to achieve consensus.

In order to achieve privacy, consensus algorithms have been proposed so that the states are masked or encrypted \cite{hadjicostis2018privary} prior to being transmitted over the network. 
In \cite{mo2016privacy}, for example, the authors propose a   privacy-preserving average consensus algorithm to guarantee the privacy of the initial  state along with guarantee of asymptotic consensus of the agents  to  the  average of the initial values, by adding and subtracting random noise to the consensus process. 
A dynamic approach is presented in \cite{altafini2019dynamical}  to achieve consensus,   while ensuring that the agents do not reveal their initial states.
In \cite{manitara2013privacy} and \cite{charalambous2019privacy}  the authors developed  distributed privacy-preserving    consensus algorithms  that  enable the agents to asymptotically reach consensus without  having  to  reveal  their initial conditions. 
These methods are designed mainly for \textit{average consensus} protocols where agents reach consensus at a value equal to the arithmetic average of their initial states.

\subsection{Motivation and Contribution}
In all existing approaches \cite{mo2016privacy,charalambous2019privacy,manitara2013privacy,altafini2019dynamical} the focus has been on the privacy of the agents' initial states. 
However, the states of the agents during the consensus process is also important and should also be protected. 
Furthermore, if the eavesdroppers are somehow able to access the parameters of the encoding process that mask the state values, they will be able to perfectly decode the states. 
In addition, previous methods do not consider communication constraints while designing  privacy-preserving protocols. 
Due to the limited bit-rate, if the transmitted messages are  quantized to meet this limited bit-rate constraint, there is no guarantee that these methods will ensure that the agents reach consensus.

In this article, we expand on previous works on adversarial consensus and consider the case where adversarial agents that do not participate in the consensus objective are able to eavesdrop into the inter-agent communications in an effort to estimate the states of the agents.
The success of the eavesdropping mechanism is assumed to be uncertain, that is, an adversary can intercept a transmission at any time with  probability less than 1. 
 This adversarial model is different from other types of adversarial models where adversaries participate in the consensus process and purposefully injects false data to affect the outcome of the consensus. 
It has been shown that in certain cases the agents can detect these malicious nodes (adversaries) in the network and disconnect such nodes from the consensus graph. 
Once the adversaries are identified and removed, they cannot further influence the consensus outcome \cite{pasqualetti2007distributed,tomasin2011consensus}.
In contrast, the agents in our eavesdropping adversarial setting cannot exclude the adversaries from eavesdropping. 
In fact, the agents might not even know whether such eavesdropping adversaries are present or not.
Therefore, the agents of the consensus network must follow a consensus protocol which ensures that the adversaries are not able to estimate their states.
That is, their true state values must be protected from being revealed to the adversaries at all times.

The protection (privacy) metric in this paper is defined in terms of the estimation error of the agents' states by the adversaries.
The objective of the agents is then to design a consensus algorithm that will ensure protection of their states against adversarial eavesdropping.
Furthermore, due to the bit-rate constraint of the underlying communication channel, the developed consensus protocol must be designed to achieve consensus even under finite data-rate constraints.
Protection in the proposed consensus protocol is achieved by not transmitting the state of the agent but rather by transmitting an \textit{innovation signal} that contains only the information required for the agents to reconstruct the state of their neighboring agents, which, however, is not useful to an eavesdropping adversary to achieve the same, unless it has perfect, uninterrupted interception capabilities.
If, at any instant of time, the adversary fails to intercept the transmitted signal, it will be incapable to reconstruct the state of the agent for any future time.

Encryption-based techniques that provide protection in the presence of adversaries have also been reported in~\cite{ruan2017secure, kishida2018encrypted}.
However, encryption-based approaches are generally computationally more expensive  and, importantly, they require additional bandwidth to implement since encryption requires a larger number of bits.
However, if desired, one can always add an extra layer of protection by further encrypting the innovation signals after quantizing them and prior to transmission.
In that context, the proposed approach can be viewed as complementary to these encryption-based approaches.

The major contributions of this work are: 
First, we show that the standard consensus protocol \cite{blondel2005convergence,zhu2010discrete,chen2013consensus,nedic2014lyapunov,fang2005information} is not \textit{protected} in the presence of adversarial agents. 
Second, we propose an innovation-communication based consensus (ICC) algorithm, and formally show that the proposed algorithm is \ep{\epsilon} in the presence of adversaries. 
That is, the mean-square error in the adversaries' estimate of the agent states is at least $\epsilon$ times the true state values (see Definition~\ref{Df:e-protected}).
We also show that the ICC algorithm does not affect the consensus value.
Third, we consider the case where only a finite number of bits is available for inter-agent communications, and derive conditions on the bit-rate to ensure consensus.
Fourth, we design dynamic quantizers to quantize the signals prior to transmission.
Dynamic quantizers can ensure that each agent reaches consensus, whereas static quantizers, in general, only ensure \textit{practical consensus}.
We show that, if the available bit-rate is greater than a certain threshold, then the bit-rate does not affect the  $\epsilon$-protection guaranteed by the ICC algorithm. 
Fifth, we show that the agents are able to reach consensus under any data-rate greater than a certain threshold, which solely depends on the parameters of the consensus network.

The rest of the paper is organized as follows: A summary background on distributed consensus is presented in Section~\ref{Sec:background}. 
The model of the adversary considered in this paper and the notion of $\epsilon$-\textit{protection} are introduced in Section~\ref{Sec:adversarial_model}.
In Section \ref{Sec:0 protection}, we formally show that the classical consensus algorithm is unprotected.
Next, in Section \ref{Sec:ICC}, we develop an innovation-based inter-agent communication scheme, and we propose a new consensus protocol to achieve $\epsilon$-\textit{protection}.
In Section~\ref{Sec:quantized consensus}, we show that the proposed consensus protocol  also guarantees consensus under a finite bit-rate constraint by utilizing suitably designed dynamic quantizers.
Simulation results are discussed in Section~\ref{Sec:simulation} that corroborate the theoretical analysis presented in this paper, and further reveal some additional desirable features of the proposed framework.
Finally, the paper is concluded in Section~\ref{Sec:Conclusion}.

 To enhance the readability of the main results, all the proofs are presented in the Appendix.



\section{Background and Preliminaries} \label{Sec:background}

\subsection{Classical Linear Consensus}

Consider a multi-agent system with $N$ agents, where the connectivity amongst the agents is dictated by a \textit{strongly connected}
digraph (directed graph) $\G(\V,\Ed)$ with $\V=\{1,\ldots, N\}$ representing the set of agents, and with $\Ed \subseteq \V\times \V$ representing the inter-agent connections such that $E_{ij}\triangleq (i,j)\in \Ed$ if and only if agent $i$ is an \textit{incoming} neighbor of agent $j$.
The incoming neighbors of agent $i$ are those agents that send information to agent $i$.
The \textit{incoming} neighbor set of agent $i$ is denoted by $\N_i \triangleq \{j~|~E_{ji}\in \Ed\} \subseteq \V$.
The \textit{outgoing} neighbor set $\N^i\triangleq \{j~|~E_{ij}\in \Ed\}$ denotes the set of agents that receive information from agent $i$. 
For an undirected graph, $E_{ij} \in \Ed$ if and only if $E_{ji} \in \Ed$ and $\N_i=\N^i$.
Throughout this paper, and unless stated otherwise, we will simply write neighbor to denote an incoming neighbor.

\begin{definition} \label{Df:outgoing links}
For agent $i$, the edges $\{E_{ij}\}_{j\in \N^i}$  and $\{E_{ji}\}_{j\in \N_i}$ denote its \textit{outgoing links}  and its \textit{incoming links}, respectively. 
Agent $i$ sends its measurements through links $E_{ij}$ and receives measurements through $E_{ji}$. 
\end{definition}
The agents follow the consensus dynamics \cite{blondel2005convergence,zhu2010discrete,chen2013consensus,nedic2014lyapunov,fang2005information}
\begin{subequations}\label{eq:consensus-protocol}
\begin{align}
     &x_{i}(k+1)=x_i(k)+ u_i(k),  \\
     &u_i(k)=\sum_{j\in \N_i}\kappa_{ij}(x_j(k)-x_i(k)),  \label{eq:control}  \\
    &x_{i}(k+1)=x_i(k)+\sum_{j\in \N_i}\kappa_{ij}(x_j(k)-x_i(k)),  \label{eq:discrete_consensus}
\end{align}
\end{subequations}
where $x_i(k) \in \R^d$ is the state of agent $i$ at time $k$. 
For simplicity of the exposition, henceforth, we consider $d=1$.
The gains $\kappa_{ij} \ge 0$ in \eqref{eq:control} are chosen  such that $\sum_{j\in \N_i}\kappa_{ij} \le 1$ holds for all $i\in \V$.
Agent $j$ shares its state $x_j$ with its outgoing neighbors $i\in \N^i$ to help agent $i$ compute the signal $u_i$ that is used to update the state $x_i$ in \eqref{eq:control}-\eqref{eq:discrete_consensus}.

The evolution of the states of all the  agents $x(k)\triangleq [x_i(k),\ldots,x_N(k)]\T$ can then be  represented as 
\begin{align} \label{eq:consensus}
    x(k+1)=Lx(k),
\end{align}
where, for all $i$ and $j$, the matrix $L$
is defined as
\begin{align*}
    [L]_{ij}=\begin{cases}
             \kappa_{ij},\qquad & j \in \N_i,\\
            1-\!\!\sum\nolimits_{j\in \N_i}\!\!\kappa_{ij} \triangleq \kappa_{ii}, & j=i,\\
             0, & \text{otherwise}.
             \end{cases}
\end{align*}
By construction, the matrix $L$ has an eigenvalue of $1$ with corresponding eigenvector $\textbf{1}\triangleq [1,\ldots, 1]\T$ \cite{xiao2007distributed}. 
In order to proceed, let us discuss certain properties of $L$ that will be used in the subsequent analysis.

\begin{theorem}[\hspace{-.1 pt}\cite{olfati2004consensus}] \label{T:prel1}
If $\G$ is strongly connected, then the eigenvalues $\{\lambda_1,\lambda_2,\ldots,\lambda_N\}$ of $L$ have the property
\begin{align*}
    1=\lambda_1 > |\lambda_2| \ge \cdots \ge |\lambda_N|~.
\end{align*}
\end{theorem}


\begin{theorem} \label{T:wl}
Let $\G$ be strongly connected.
Then, there exists  a matrix $L_1$ and a vector $w_\ell$ such that the matrix $L$ can be written as follows
\begin{align*}
    L=\textbf{1}w_\ell\T+L_1, 
\end{align*}
where $w_\ell\T\textbf{1}=1$ and $L_1\textbf{1}=w_\ell\T L_1=0$.
\end{theorem}
\begin{proof}
See Appendix \ref{AP:wl} for the proof.
\end{proof}
The vector $w_\ell$ in Theorem \ref{T:wl} is the left eigenvector of $L$ corresponding to the eigenvalue 1.
Define now the scalar $c(k)=w_\ell\T x(k)$.
As a consequence of Theorem \ref{T:wl}, we have that $c(k+1)=w_\ell\T x(k+1)=w_\ell\T L x(k)=w_\ell\T x(k)=c(k)$. 
Therefore, $c(k)$ is a conserved quantity under the consensus protocol \eqref{eq:discrete_consensus}, and the agents reach the final consensus value 
 \begin{align} \label{eq:x_final_value}
     \lim_{k\to \infty} x(k) \triangleq x_\infty =\textbf{1}c(0)= \textbf{1}(w_\ell\T x(0)).
 \end{align}

Let us define the matrix $J \triangleq \textbf{1}w_\ell\T$.
Certain properties of the matrix $J$ that will be useful in this paper are summarized in the following proposition.
\begin{pro} \label{Pro:prel3}
For all $k\in \bbN$,
\begin{enumerate}[(i)]
    \item $J$ is idempotent. 
    \item $L_1J=JL_1=0$ and $LJ=J$.
    \item $(L-J)^k=L^k-J^k$.
    \item $(L-I)L^k=(L-I)(L-J)^k$.
    \item $\rho(L-J)=|\lambda_2| <1$, where $\rho(\cdot)$ denotes the spectral radius.
    \item $\lim_{k\to \infty}  (L-J)^k =0 $ and $\lim_{k \to \infty } L^k = J$.
\end{enumerate}
\end{pro}
\begin{proof}
Please see Appendix \ref{AP:Pro_prel3} for the proof.
\end{proof}

 \begin{corr}
 If $L$ is a doubly stochastic matrix, i.e., all the elements of $L$ are non-negative and each row and each column sums to $1$, then  $w_\ell=({1}/{N})\textbf{1}$,  $J=({1}/{N})\textbf{1}\textbf{1}\T$, and $x_\infty=\sum_{i=1}^N x_i(0)/N$.
 \end{corr}
 
 \subsection{Consensus with Quantized Measurements} \label{sec:quant_prel}
 
The agents participating in consensus communicate over a shared channel with a bit-rate of $\mathcal{B}_0$ bits per time-step.
Therefore, the signals need to be encoded (quantized) prior to transmission to meet this bit-rate constraint.
The available bit-rate $\mathcal{B}_0$ is shared by all the agents. 
The total number of messages sent over the network at any time instant  is $\N\triangleq \sum_{i=1}^N|\N_i|$.
If each encoding requires $b_0$ overhead bits, then each message can be encoded into $b\triangleq \lfloor{\mathcal{B}_0} \slash {\N}\rfloor-b_0$ bits.
One of the objectives of this paper is to derive sufficient conditions on $b$ to ensure protected consensus.

In the consensus protocol \eqref{eq:consensus-protocol}, the states of all the agents are communicated, and hence, all prior works on quantized consensus \cite{frasca2008average,zhang2015periodic,aysal2008distributed,thanou2012distributed,li2013unified} solely focus on the case where the states of the agents are quantized. 
These works ensure that (practical) consensus  can be achieved if the quantization process satisfies certain conditions.
However, there is no formal guarantee that consensus will be achieved if each agent quantizes and shares a different signal other than its own state.
We will show later on that, in order to retain protection against adversarial eavesdropping, a different signal ($\xi$) than the state ($x$) has to be quantized and transmitted.
We therefore need to design a new quantization scheme under which consensus can be achieved even though each agent does not quantize and transmit its own state. 
Later in this paper, we will follow an approach similar to \cite{thanou2012distributed} to design dynamic quantizers  to achieve quantized consensus with asymptotic protection against adversarial eavesdropping.
By a careful design of the dynamic quantizer parameters, the authors in \cite{thanou2012distributed} show that \textit{all} states reach consensus.
However, in our problem we cannot directly apply the dynamic quantizers of \cite{thanou2012distributed} since those quantizers were designed to quantize the states that follow the consensus protocol \eqref{eq:consensus} on a balanced graph $\G$ (i.e., $L$ is a doubly stochastic matrix).
The dynamic quantizers of \cite{thanou2012distributed} therefore do not provide any guarantee whether consensus will be achieved if a different signal other than the state itself is quantized for a general digraph $\G$.
In Section~\ref{Sec:quantized consensus} we show how the approach in \cite{thanou2012distributed} can be modified to achieve both consensus and adversarial protection.

\section{Consensus in the  Presence of Adversaries} \label{Sec:adversarial_model}
We consider the scenario where the agents perform consensus in a compromised communication network and inter-agent communications are intercepted by adversarial agents not participating in the consensus. 
We further assume that the adversaries do not alter the transmitted messages over the communication links. 
Instead, their primary objective is to estimate the states of the agents by intercepting the transmitted signals. 
Such adversarial eavesdropping \cite{anand2005quantifying, zhang2010p} is common in military and defense applications where the agents decide a rendezvous point/trajectory by using a consensus algorithm. 
If the adversaries have a perfect estimate of the agents' states then they can ambush the agents before the agents could achieve their goal.
Since adversaries do not alter the transmitted messages, the agents are not able to detect the presence of eavesdropping adversaries.
In contrast, when the adversaries actively take part in the consensus by injecting false data to corrupt the consensus outcomes, the agents can detect such malicious intents and remove the adversaries from the consensus process; see for example \cite{pasqualetti2007distributed, tomasin2011consensus}. 
In the presence of eavesdropping adversaries, the agents cannot detect the presence of such adversaries, and hence, they can not change the consensus graph (e.g., by removing inter-agent compromised links) to prohibit the adversaries from eavesdropping. 
As a result, protection against such passive eavesdropping adversaries 
 becomes more challenging.


\subsection{Adversarial Eavesdropping Model}
For the simplicity of this exposition, we consider a scenario where the number of adversaries is the same as the number of agents. 
Each adversary is assigned to estimate the state of one  agent at all times.
Without loss of generality, we assign adversary $i$ to agent $i$, i.e., adversary $i$ will estimate the state of agent $i$; such an estimate at time $k$ will be denoted as $\xad_i(k)$.
The adversaries do not know the initial states of the agents. 
The only way therefore for adversary $i$ to estimate the state $x_i$ of agent $i$ is by tapping the outgoing communication links of agent $i$.
The outgoing set $\N^i$ contains all the agents that receive measurements from agent $i$. 
Adversary $i$ then eavesdrops the outgoing links $E_{ij}$, $j\in \N^i$ to intercept the messages sent by agent $i$.

The eavesdropping mechanism is probabilistic and may lead to failed interception of the message, similar to the model in \cite{tsiamis2019state}.
Eavesdropping success is modeled by a Bernoulli random variable.
Associated with adversary $i$, let $\mu_i(k) \in \{0,1\}$ be the random variable representing the event of whether eavesdropping at time $k$ was successful for adversary $i$ or not, i.e.,
\begin{align} \label{eq:mu_bernoulli}
    \mu_i(k)=\begin{cases} 1,\!\! &\text{successful eavesdropping of agent $i$ at time $k$},\\
    0, &\text{otherwise}.
    \end{cases}
\end{align}
  The sequence of random variables $\{\mu_i(k)\}_{k\in \bbN_0}$ is {i.i.d}, with $\bbP(\mu_i(k)=1)=\gamma_i=1-\bbP(\mu_i(k)=0)$, for some $\gamma_i\in (0,1)$. 
  For simplicity of the analysis, we consider $\gamma_i = \gamma$ for all $i \in \V$.

\subsection{$\epsilon$-Protection Against Imperfect Adversaries}

Let $\hat{x}^{\rm ad}(k) =[\xad_i(k),\ldots,\xad_N(k)]\T$ denote the collective estimate of the state $x(k)$ of all agents by the adversaries at time $k$ and let $e^{\rm ad}(k)\triangleq x(k)-\hat{x}^{\rm ad}(k)$ be the estimation error. 
It is noteworthy that at any time $k$ the estimated state $\xad(k)$ and  the estimation error $\ead(k)$ are random variables due to the probabilistic nature of the eavesdropping mechanism.

Since we are interested in protecting the agent state values from being perfectly estimated by the adversaries, we need to define a suitable metric to quantify such privacy.
The following definition provides a metric for state-protection (or state-privacy) that is adopted in this article.
\begin{definition} \label{Df:e-protected}
A consensus algorithm is $\epsilon$-\textit{protected} against the adversaries for an initial state $x(0)\in \R^{N}$, if there exists  $\epsilon>0$ and $c>0$ such that, for all $k\in \bbN_0$,
\begin{align} \label{eq:df_ep}
    \E[\ead(k)\T\ead(k)]\ge \epsilon\|x(k)\|^2 +c.
\end{align}
The consensus algorithm is \textit{asymptotically} $\epsilon$-\textit{protected} if 
\begin{align} \label{eq:df_aep}
    \liminf_{k\to \infty}\E[\ead(k)\T\ead(k)]\ge \epsilon\|x_\infty\|^2 +c.
\end{align}
The algorithm is called  $0$-\textit{protected} (\textit{asymptotically $0$-protected})  if there does not exist any $\epsilon>0$ such that  \eqref{eq:df_ep} (respectively \eqref{eq:df_aep}) holds.
\end{definition}
The constant $c$ in Definition \ref{Df:e-protected} guarantees that, for a protected consensus algorithm, $\E[\ead(k)\T\ead(k)]>0$ even when $x(k) =0$, and similarly, for an asymptotically protected algorithm, $\liminf_{k\to \infty}\E[\ead(k)\T\ead(k)] >0$ even when $x_\infty = 0$.
\begin{rem} \label{rem:eps-protection}
Based on Definition \ref{Df:e-protected}, we notice that \textit{asymptotic $\epsilon$-protection} is a necessary condition for $\epsilon$-\textit{protection}.
Therefore, if the agents are prevented from achieving asymptotic $\epsilon$-protection, then they cannot achieve $\epsilon$-protection.
Thus, it is sufficient for the adversaries to ensure that the agents cannot achieve asymptotic $\epsilon$-protection, for any $\epsilon>0$, in order to make the consensus protocol unprotected, i.e., $0$-\textit{protected}.
\end{rem}

\subsection{Imperfect and Partial Knowledge of the Consensus Model}

It is assumed that the adversaries know the communication protocol that the agents are using.
In addition, we assume that  adversary $i$ 
does not know the underlying inter-agent communication structure or the parameters $\{\kappa_{ij}\}_{j\in \N_i}$.
In many applications, the consensus matrix $L$ is time varying and randomly chosen (e.g., push-sum algorithms, gossip algorithms \cite{chen2020privacy}), and hence, the adversaries cannot estimate $L$.
The scenario considered in this paper is an instance of \textit{symmetric partial knowledge}, in the sense that all adversaries have the same information. 
The scenario where  adversaries have asymmetric knowledge of the consensus model (e.g., some adversaries know more elements of the $L$ matrix than other adversaries) is also of interest, but it is beyond the scope of this work.
\section{Classical Consensus is $0$-Protected} \label{Sec:0 protection}

We first show that the classical consensus protocol, where each agent shares its state $x_i(k)$,  is not protected from  eavesdropping. 
In particular, the adversaries are able to design an algorithm to construct an estimate $\xad(k)$ so  that the consensus protocol in \eqref{eq:consensus} is $0$-\textit{protected}.
In order to see this, let adversary $i$ follow the estimation dynamics 
\begin{equation}\label{eq:adver_dynamics_i}
\begin{aligned} 
    &\xad_i(k)=\mu_i(k)x_i(k)+(1-\mu_i(k))\xad_i(k-1),\\
    &\xad_i(-1)=0.
\end{aligned}    
\end{equation}
By following \eqref{eq:adver_dynamics_i}, each adversary updates its estimate only when it has successfully eavesdropped its assigned agent (i.e., $\mu_i(k)=1$), otherwise it retains the previous estimate.
The main result of this section is presented in the following theorem.
\begin{theorem} \label{T:0-protection}
The consensus protocol \eqref{eq:consensus} is  $0$\textit{-protected}.
\end{theorem}
\begin{proof}
The proof is presented in Appendix \ref{AP:0-protection}.
\end{proof}

Intuitively, it makes sense that the classical consensus algorithm will be \ep{0} for the following reason.
We have that $\bbP(\max_{t }{\mu_i(t)} = 0) =0$, or in other words, there exists a future time when eavesdropping will be successful almost surely. 
Since $x(k) \to x_\infty$, successful interception of a transmission late enough will reveal the consensus value to the adversary.

The  estimation dynamics \eqref{eq:adver_dynamics_i} of adversary $i$, which does not require knowledge of the network parameters $\kappa_{ij}$, ensures that the classical consensus protocol is not \aep{\epsilon} for any $\epsilon >0$, and hence it is $0$-\textit{protected} (Remark~\ref{rem:eps-protection}).
Since the estimation dynamics \eqref{eq:adver_dynamics_i} depend neither on the network parameters nor on the dynamics \eqref{eq:consensus}, we obtain the following corollary from Theorem~\ref{T:0-protection}.

\begin{corr} \label{corr:0-potection}
Any consensus protocol requiring that the agents share their actual state is \ep{0}.
\end{corr}

In contrast to Theorem \ref{T:0-protection}, which shows that the classical consensus protocol \eqref{eq:consensus} is \ep{0}, Corollary~\ref{corr:0-potection} states a much stronger result, namely,  that \textit{any} consensus protocol with true state communication is also \ep{0}.
Corollary \ref{corr:0-potection} motivates the search for new consensus algorithms that do not require that the agents share their actual state in order to achieve $\epsilon$-\textit{protection}.

\section{Innovation-Based Communication Scheme} \label{Sec:ICC}

In the classical consensus protocol \eqref{eq:consensus-protocol}, the purpose of sharing  the state $x_j(k)$ is to help agent $i$ compute its control input $u_i$, since $u_i$ contains the term $\kappa_{ij}x_j(k)$ in \eqref{eq:control}. 
This sharing of the true state over the compromised link makes the system $0$-\textit{protected} against adversaries, as discussed in the previous section.
Therefore, in hoping to achieve  $\epsilon$-\textit{protection}  for some $\epsilon>0$, the agents must not send their true state information.
However, in order to follow the consensus dynamics \eqref{eq:discrete_consensus}, agent $i$ must have knowledge of $x_j(k)$ for all $j\in \N_i$.
In order to achieve both these objectives (consensus and $\epsilon$-\textit{protection}), we propose the innovation-communication consensus (ICC) protocol in Algorithm~\ref{AL:ICC} that prescribes  agent $i$ to communicate an \textit{innovation} signal $\xi_i$ instead of its actual state $x_i$.

The innovation signal $\xi_i$ in Algorithm~\ref{AL:ICC} is very similar to the signal $u_i$ in \eqref{eq:control} except for the fact that $u_i$ requires the true state value $x_j$ and $\xi_i$ uses an estimate $\hat{x}_j^i$.
Since the agents do not receive the true state values from their neighbors, they need to locally estimate their neighbors' states.
In Algorithm~\ref{AL:ICC}, $\hat x_j^i(k)$ denotes the estimate of agent $j$'s state  by agent $i$.
However, in order to estimate the neighbors' states, agent $i$ does not require knowledge of agent $j$'s local parameters (e.g., $\{\kappa_{j\ell}\}_{\ell \in N_j}$).
The sequence of events that take place locally for agent $i$ from time-step $k$ to time-step $k+1$ is as follows: 
\begin{align} \label{eq:event_seq}
    &\cdots\to\text{send }\xi_i(k\!-\!1),\text{ receive } \{\xi_j(k\!-\!1)\}_{j\in \N_i}\!\to \nonumber \\ & \text{update}  \text{ the estimate } 
   \hat{x}_j^i(k) \to \text{compute } \xi_i(k)\to \text{ update}  \\ \nonumber
   &\text{the state }   x_i(k\!+\!1) \to \text{send } \xi_i(k), \text{ receive } \{\xi_j(k)\}_{j\in \N_i}\to \cdots
\end{align}
\begin{algorithm}
\caption{Innovation-Communication Consensus (ICC)} \label{AL:ICC}
\begin{algorithmic}[1]
\STATE Define {$\small {\xi_i(-1)}=x_i(0)$} for all $i\in \V$\\
\FOR{$k\ge 0$}
\FORALL{$i=1,2,\ldots, N$}
\STATE agent $i$ sends ${\xi_i(k-1)}$ to $j\in \N^i$, 
\FORALL{$j\in \N_i$}
\STATE $\hat x_j^i(k)=\hat x_j^i(k-1)+{\xi_j(k-1)},\quad \hat x_j^i(-1)=0$,
\ENDFOR
\STATE ${\xi_i(k)}=\sum_{j\in  \N_i}\kappa_{ij}(\hat x_j^i(k)-x_i(k))$
\STATE $x_i(k+1)=x_i(k)+{\xi_i(k)}$
\ENDFOR
\ENDFOR
\end{algorithmic}
\end{algorithm}

The following two theorems characterize the key properties of the ICC algorithm. 
Theorem \ref{T:ICC2} ensures that the consensus value of the ICC algorithm is the same as the classical consensus protocol of \eqref{eq:consensus-protocol}, and Theorem \ref{T:ICC} ensures that the ICC consensus protocol is $\epsilon$-\textit{protected}.

\begin{theorem} \label{T:ICC2}
By following the ICC algorithm, $x(k)$ converges to $x_\infty$ given in \eqref{eq:x_final_value}.
\end{theorem}
\begin{proof}
The proof is presented in Appendix \ref{Ap:ICC2}.
\end{proof}

From the proof of Theorem~\ref{T:ICC2} (in particular, from equations \eqref{eq:hat_x_j_dynamics}, \eqref{eq:x_j_dynamics} and the subsequent discussion) 
we obtain that $\hat{x}^i_j(k)=x_j(k)$ for all $i\in \V$ and for all $j\in \N_i$,  where $\hat x^i_j(k)$ is defined in line~6 of Algorithm \ref{AL:ICC}.
Consequently, it follows from line~8 of Algorithm~\ref{AL:ICC}  that $\xi_i(k)=\sum_{j\in \N_i}\kappa_{ij}(x_j(k)-x_i(k))$. 
Line 9 of the algorithm can then be rewritten as $x_i(k+1)=x_i(k)+\sum_{j\in \N_i}\kappa_{ij}(x_j(k)-x_i(k))$, which is the same as equation \eqref{eq:discrete_consensus}.
Therefore, from the agents' perspective, the ICC algorithm does not alter the consensus dynamics, and the agents still follow \eqref{eq:discrete_consensus} even though they are communicating via the signals $\xi_i(k)$ which are not the agents' true state values. 

In Section~\ref{Sec:0 protection} (discussion after Theorem~\ref{T:0-protection}) we notice that the classical consensus protocol is \ep{0} due to the possibility that a late interception of a transmission reveals the consensus value, as the state converges exponentially.
However, from the construction of Algorithm~\ref{AL:ICC}, we notice that $\xi_i(k) \to 0$ as $k$ increases.
Thus, late interception of a transmission does not reveal any further information about $x_i(k)$ since $\xi_i(k)$ becomes less and less correlated to $x_i(k)$ as $k$ increases.

\begin{theorem} \label{T:ICC}
The ICC algorithm is \ep{(1-\gamma)^2}.
\end{theorem}
\begin{proof}
The proof is presented in Appendix \ref{Ap:ICC}.
\end{proof}

We conclude this section with the following remark.

\begin{rem}
The construction of the ICC algorithm is based on the idea that an error incurred due to failed eavesdropping cannot be compensated entirely by successful eavesdropping in the future.  
The amount of protection obtained by the ICC   algorithm does not depend on the network structure or the initial state $x(0)$, rather, it solely depends on the eavesdropping success parameter $\gamma$. 
As expected, the amount of protection decreases as $\gamma$ increases.
\end{rem}
\section{Consensus with Limited Data-Rate} \label{Sec:quantized consensus}

Although quantized consensus is a well-studied problem in the literature \cite{frasca2008average,zhang2015periodic, aysal2008distributed,thanou2012distributed}, in all those studies the states $x_i(k)$ themselves are quantized and transmitted.
As shown in Section~\ref{Sec:ICC}, in order to ensure protection against adversaries, instead of the state $x_i(k)$ the signal $\xi_i(k)$ is transmitted. 
As a result, agent $i$ would receive a quantized version of $\xi_j(k)$ from its neighbors $j\in \N_i$. 
We therefore need to ensure that the agents reach consensus while only using the quantized version of $\xi_i(k)$.
However, not all quantization schemes that ensure satisfaction of the data-rate constraint can also ensure consensus (and stability) \cite{thanou2012distributed}.
Therefore, it is important to independently verify  the existence of a quantizer that ensures consensus and also satisfies the data-rate constraint.
With this objective in mind, next we design  (dynamic) quantizers for the agents to achieve consensus using the quantized transmitted signals $\xi_i(k)$.

\subsection{Dynamic Fixed-Length Quantizers} \label{Sec:quant}
A quantizer $q:(x_{\min},x_{\max})\subseteq \R \to \{0,\ldots, 2^b-1\}$ is a function that maps any $x\in (x_{\min},x_{\max})$ to a $b$-bit binary word.
For example, a uniform quantizer takes the form 
\begin{align*}
    q(x)=\left\lfloor\frac{x-x_{\min}}{\Delta} \right\rfloor,
\end{align*}
where $\Delta=(x_{\max}-x_{\min})/{2^b}$.
Although $q$ is defined on the interval $(x_{\min},x_{\max})$, one can extend the definition of $q$ on the entire real line as follows
\begin{align*}
    \tilde q(x)=\begin{cases}
    q(x),\quad &x\in (x_{\min},x_{\max}),\\
    0, & x\le x_{\min},\\
    2^b-1, & x\ge  x_{\max},
    \end{cases}
\end{align*}
where $\tilde q$ is the extended version of $q$. 
The region outside the interval $[x_{\min},x_{\max}]$ is the \textit{saturation} region of the quantizer $\tilde q$.

For a dynamic $b$-bit quantizer, the boundaries $x_{\min}$ and $x_{\max}$ are not fixed and are given as extra inputs.
That is, a dynamic $b$-bit quantizer is a mapping  of the form $q_d: \R\times \R \times \R_+ \to  \{0,\ldots, 2^b-1\}$ such that 
\begin{align*}
    &q_d(x,\alpha,\beta)=\begin{cases}
   \left\lfloor \displaystyle\frac{x - \alpha}{\Delta} \right\rfloor,\quad &x\in (\alpha,\alpha+\beta),\\
    0, & x\le \alpha,\\
    2^b-1, & x\ge \alpha+\beta,
    \end{cases}
\end{align*}
where $\Delta=\beta/2^{b}$.
The dynamic nature of the quantizer emerges when the parameters $\alpha$ and/or $\beta$ are varied over time.
By construction, such quantizers are defined over the whole real line $\R$ with  saturation region $(-\infty,\alpha]\cup[\alpha+\beta,\infty)$. 

While $q_d$ encodes a real-valued signal to a $b$-bit binary word, the decoding of the quantized signal is performed as follows:
\begin{align} \label{eq:decoding}
    {x}^q=q_d(x,\alpha,\beta)\Delta+\frac{\Delta}{2}+\alpha,
\end{align}
where $x^q$ is the decoded version of $x$ after being quantized by $q_d$. 
To simplify the terminology in the subsequent sections, we will often refer to $x^q$ as \textit{quantized $x$} for any signal $x$.
Let $\Delta^q(x)\triangleq x - x^q$ be the quantization error associated with quantizer $q_d$.
One then observes that for any $x\in (\alpha,\alpha+\beta)$
\begin{align} \label{eq:quant_error}
    |\Delta^q(x)|=|x-x^q|\le \frac{\Delta}{2}=\frac{\beta}{2^{b+1}}.
\end{align}

 Dynamic quantizers for higher dimensions can be defined accordingly. 
 For instance, for any $x,\alpha\in \R^n$ and $\beta\in \R^n_+$, define 
$q_d(x,\alpha,\beta)=[q_d(x_1,\alpha_1,\beta_1),q_d(x_2,\alpha_2,\beta_2), $ $~ \ldots,  q_d(x_n,\alpha_n,\beta_n)]\T$ to be an $n$-dimensional vector-valued quantizer.
Similarly, decoding of an $n$-dimensional quantized vector is performed element-wise using \eqref{eq:decoding}.
The quantization error vector $\Delta^q(x)\triangleq [\Delta^q(x_1),\ldots,\Delta^q(x_n)]\T$ is defined accordingly.
We will only consider dynamic quantizers for the subsequent sections and will henceforth suppress the subscript $d$ by simply writing $q$ for brevity.

\subsection{Design of a Dynamic Quantizer} \label{Sec:dyn_quant}
Recall that Algorithm \ref{AL:ICC} (line 4) requires agent $i$ to sent $\xi_i(k-1)$ at time $k$ to its outgoing neighbors $\N^i$.
In this subsection, we design dynamic quantizers for the agents to quantize their signals $\xi_i(k-1)$ at time $k$ prior to transmission.
Without loss of generality, let us assume that there exists $x_{\max} > x_{\min}$ such that $x_i(0) \in (x_{\min},x_{\max})$ for all $i\in \V$.
In order to send $\xi_i(k-1)$ at time step $k$, agent $i$ uses a dynamic quantizer of the form $q_i(\cdot,k) \triangleq q(\cdot, \alpha_i(k),\beta(k))$ as follows 
\begin{align}
    &\alpha_i(k)=\xi_i^q(k-2)-\frac{\beta(k)}{2}, \label{eq:alpha_dynamics}\\
   & \frac{\beta(k+1)}{2}=\frac{3\sqrt{N} \beta(k)}{2^{b+1}}
    +\frac{4\sqrt{N}}{2^{b+1}}\sum_{\ell=0}^{k-1}|\lambda_2|^{k-1-\ell}\beta(\ell)\label{eq:beta_dynamics} \\
    &\quad ~~~~~~~~~~~~+4|\lambda_2|^{k-1}\sqrt{N}\max\{|x_{\min}|,|x_{\max}|\}, \nonumber\\
    & \alpha_i(0)=x_{\min}, ~~\beta(0)= (x_{\max}-x_{\min}),\label{eq:alpha_initial}\\
   &\beta(1)=3\max\{|x_{\min}|,|x_{\max}|\} \label{eq:beta_initial},
\end{align}
where $|\lambda_2|$ is the absolute value of the second largest eigenvalue of matrix $L$ as denoted in Theorem \ref{T:prel1}. 
Let $\Delta_i^q(\cdot,k)$ be the quantization error associated with the quantizer $q_i(\cdot,k)$ at time $k$. 
Then from \eqref{eq:quant_error}, we have
\begin{align} \label{eq:delta_ik}
    |\Delta^q_i(\xi,k)|\le \beta(k)/2^{b+1},
\end{align}
for all $\xi \in [\alpha_i(k),\alpha_i(k)+\beta(k)]$.

As observed in \eqref{eq:decoding}, in order to decode the output of the quantizer $q(x,\alpha,\beta)$, the receiver needs the values of the parameters $\alpha$ and $\beta$.
Therefore, if agent $j$ receives a quantized measurement from agent $i$ that uses $q_i(\cdot,k)$ to quantize its signal, then agent $j$ needs to know $\alpha_i(k)$ and $\beta(k)$ in order to decode the quantized version of $\xi_i$.
From the expression of $\beta(k)$ in \eqref{eq:beta_dynamics}, notice that all agents can precompute $\beta(k)$ as long as they have knowledge of $\lambda_2$ and $N$.
A slightly conservative value of $\beta(k)$ may be derived that does not require knowledge of the network parameter $\lambda_2$.
Here we assume that the agents  have computed the values of $\beta(k)$'s prior to  starting the consensus algorithm.

Recall from \eqref{eq:event_seq} that agent $i$ computes  $\xi_i(k)$ during the interval $[k,k+1]$ and at  time-step $k+1$ agent $i$  quantizes $\xi_i(k)$ using the quantizer $q_i(\cdot,k+1)$ with the corresponding $\alpha_i(k+1)=\xi_i^q(k-1)-\beta(k+1)/2$.
Agent $i$ then forwards the quantized measurement $q_i(\xi_i(k),k+1)$ to agent $j$ via the link $E_{ij}$.
During the time interval $[k,k+1]$ agent $j$ received the quantized version of $\xi_i(k-1)$  from agent $i$  and the decoded signal $\xi^q_i(k-1)$ is now available to agent $j$ at time $k+1$.
Therefore, at time $k+1$ agent $j$ is able to compute $\alpha_i(k+1)$ since it knows both $\xi_i^q(k-1)$ and $\beta(k+1)$.
 Upon receiving a quantized measurement $q_i(\xi_i(k),k+1)$ at time $k+1$, agent $j$ can use the decoder defined in \eqref{eq:decoding} to decode the received quantized measurement.

In the subsequent sections we show that the use of a dynamic quantizer \eqref{eq:alpha_dynamics}-\eqref{eq:beta_initial} indeed leads to consensus.
Before showing this fact, we provide the following lemma that characterizes certain useful properties of $\beta(k)$.
\begin{lemma} \label{L:rate_bound}
If the $b$-bit dynamic quantizer \eqref{eq:alpha_dynamics}-\eqref{eq:beta_initial} is used to encode each transmission, then $\beta(k)\to 0$ exponentially, assuming  $b$ is selected as follows
\begin{align} \label{eq:data_rate}
    b> \frac{1}{2}\log_2 N+\log_2{\left(\frac{7-3|\lambda_2|}{1-|\lambda_2|}\right)}.
\end{align}
\end{lemma}
\begin{proof}
The proof is presented in Appendix \ref{AP:L_rate_bound}.
\end{proof}
For the dynamic quantizer $q_i(\cdot,k)$ the parameter $\beta(k)$ represents the quantization range and is directly associated with the quantization error as per equation \eqref{eq:delta_ik}.
Therefore, as $\beta(k)$ goes to zero the quantization error goes to zero as well.
Lemma \ref{L:rate_bound} provides a sufficient condition for the bit-rate $b$ to ensure that the quantization error converges to zero.
While Lemma \ref{L:rate_bound} ensures that $\beta(k)$ converges to zero,  it however does not comment on the rate of convergence. 
The following lemma characterizes the rate ($\eta$) at which $\beta(k)$ converges to zero.
\begin{lemma} \label{L:beta_rate}
Let $1> \eta >|\lambda_2|$. 
If the number of bits used in the quantization scheme satisfies
\begin{align} \label{eq:b_eta}
    b= \frac{1}{2}\log_2 N+\log_2{\left(\frac{4+3(\eta-|\lambda|_2)}{\eta(\eta-|\lambda|_2)}\right)},
\end{align}
then 
\begin{align} \label{eq:beta_eta}
\beta(k) \le \bar{\beta}\eta^k,    
\end{align}
where $\bar{\beta}$ is a constant that depends on $\beta(0)$ and $\beta(1)$.
\end{lemma}
\begin{proof}
Please see Appendix \ref{Ap:L_beta_rate} for the proof.
\end{proof}

\subsection{Protected Consensus with Dynamic Quantizers}
Let  $x_0 \triangleq [x_{10},x_{20},\cdots,x_{N0}]\T$
denote the initial states of the agents.
Under quantized consensus, let each agent follow the dynamics
\begin{subequations}
\begin{align}
    & x_i(k+1)=x_i(k)+\xi_i^q(k), \quad x_i(0) =x_{i0}^q\label{eq:dyn_quant} \\
    & \xi_i(k)=\sum_{j\in\N_i}\kappa_{ij}(\hat{x}_j^i(k) -x_i(k)),\quad \xi_i(-1)=x_{i0}, \label{eq:xi_quant}\\
      & \hat x_j^i(k)=\hat x_j^i(k\!-\!1)\!+\!{\xi_j^q(k\!-\!1)},\quad \hat x_j^i(-1)\!=\!0, ~ j\in \N_i, \label{eq:local_esti_quant}
\end{align}
\end{subequations}
where $x_{i0}^q$ and $\xi_i^q(k)$ represent quantized $x_{i0}$ and $\xi_i(k)$, respectively.
Equations \eqref{eq:dyn_quant}-\eqref{eq:local_esti_quant} are similar to the ones proposed in Algorithm~\ref{AL:ICC}, except that now each agent uses the quantized signal  $\xi_i^q(k)$ instead of $\xi_i(k)$ and $x_{i0}^q$ instead of $x_{i0}$.
Note that, in absence of quantization we have that $x_{i0}^q = x_{i0}$ and $\xi^q_i(k)=\xi_i(k)$ for all $k$, and hence, the equations \eqref{eq:dyn_quant}-\eqref{eq:local_esti_quant} become exactly the same as the ones in Algorithm~\ref{AL:ICC}.

Although agent $i$ may have access to $x_{i0}$ and $\xi_i$, it still uses the quantized values $x_{i0}^q$ and $\xi_i^q$ in its dynamics \eqref{eq:dyn_quant}. 
This is to ensure that consensus is achieved. 
Surprisingly, using the true values $x_{i0}$ and $\xi_i$ in \eqref{eq:dyn_quant} may not lead to a consensus.
A direct consequence of every agent using quantized values in \eqref{eq:dyn_quant} is that agent $i$ can perfectly estimate agent $j$'s state, i.e. $\hat x_j^i(k) = x_j(k)$ for all $k\in \bbN_0$ and $j \in \N_i$.
To see this, notice from equation \eqref{eq:dyn_quant} that agent $j$'s dynamics  follow
\begin{align}
    x_j(k+1)=x_j(k)+\xi_j^q(k), \quad x_j(0) = x_{j0}^q. \label{eq:agent_j_dynamics_quant}
\end{align}
Comparing agent $j$'s dynamics \eqref{eq:agent_j_dynamics_quant} with \eqref{eq:local_esti_quant}, we obtain that $\hat{x}^i_j(k) = x_j(k)$ for all $k=\bbN_0$ and $i \in \V$ and $j \in \N_i$.
Thus, by replacing $\hat{x}_j^i(k)$ with ${x}_j(k)$, we may re-write   \eqref{eq:dyn_quant}-\eqref{eq:local_esti_quant}   as follows
\begin{subequations} \label{eq:x_xi_xiq}
\begin{align}
   & x_i(k+1)=x_i(k)+\xi_i^q(k),\quad x_i(0) =x_{i0}^q,\label{eq:dyn_quant_2}\\
   & \xi_i(k)=\sum_{j\in\N_i}\kappa_{ij}({x}_j(k)-x_i(k)), \quad \xi_i(-1)=x_{i0}, \label{eq:xi_quant_2}
\end{align}
\end{subequations}
Before proceeding further, we present the following lemma that will be paramount in the subsequent analysis.
\begin{lemma} \label{L:no_saturation}
For all $k\in \bbN_0$, $\xi_i(k-1) \in [\alpha_i(k),\alpha_i(k)+\beta(k)]$.
\end{lemma}
\begin{proof}
See Appendix \ref{Ap:L_xi} for the proof.
\end{proof}

\begin{rem} \label{rem:no_saturation}
A direct consequence of Lemma \ref{L:no_saturation} is that the signal $\xi_i(k-1)$ does not lie in the saturation region of quantizer $q_i(\cdot,k)$ and hence, when quantized using $q_i(\cdot,k)$, it obeys the bound
\begin{align} \label{eq:delta_bound}
    |\Delta_i^q(\xi_i(k-1),k)| \le \frac{\beta(k)}{2^{b+1}}.
\end{align}
\end{rem}

By substituting the expression of $\alpha_i(k)$ from \eqref{eq:alpha_dynamics},   we  have from Lemma \ref{L:no_saturation} that $|\xi_i(k)-\xi_i^q(k-1)|\le {\beta(k+1)}/{2}$.
If the number of quantization bits $b$ satisfies \eqref{eq:data_rate} (or \eqref{eq:b_eta}), then, as a result of Lemma \ref{L:rate_bound} (or Lemma~\ref{L:beta_rate}) and Lemma \ref{L:no_saturation}, we have 
\begin{align} \label{eq:xi_q_converge}
    \lim_{k\to \infty} |\xi_i(k)-\xi_i^q(k-1)|\le\lim_{k\to \infty} \frac{\beta(k+1)}{2} =0.
\end{align}
This shows that $\xi(k)$ converges to $\xi^q(k-1)$.
Furthermore, we also obtain
\begin{align*}
    |\xi_i(k)-\xi_i(k-1)|&\le \\
    |\xi_i(k)-\xi_i^q(k&-1)| + |\xi_i(k-1)-\xi_i^q(k-1)|\\
    &\le \frac{\beta(k+1)}{2}+|\Delta^q_i(\xi_i(k-1),k)|\\
    &=\frac{\beta(k+1)}{2}+\frac{\beta(k)}{2^{b+1}},
\end{align*}
and, consequently, it follows that
\begin{align} \label{eq:xi_converge}
    \lim_{k\to \infty} |\xi_i(k)-\xi_i(k-1)| =0.
\end{align}
Based on \eqref{eq:xi_converge} and \eqref{eq:xi_q_converge}, we conclude that the signals $\xi(k)$ and $\xi^q(k)$ converge.
The following theorem shows that, by using dynamic quantizers, the states $x(k)$ converge, but not necessarily to the value $x_\infty$  given in \eqref{eq:x_final_value}. 
The theorem also characterizes the derivation of the consensus value  from $x_\infty$ due to quantization.
\begin{theorem} \label{T:main}
For the dynamic quantizers designed in \eqref{eq:alpha_dynamics}-\eqref{eq:beta_initial}, assume that the number of bits $b$ satisfies \eqref{eq:b_eta} for some $\eta \in (|\lambda_2|,1) $. 
Then there exists $c\in \R$ such that
\begin{align}
   & \lim_{k\to \infty} x_k = \Tilde{x}_\infty=\textbf{1}c.
\end{align}
Furthermore,
\begin{align}
    \|\tilde{x}_\infty-x_\infty\|\le \left(\beta(0)+\frac{\bar\beta\eta}{1-\eta}\right) \sqrt{N}2^{-(b+1)}, \label{eq:consensus_deviation}
\end{align}
where $\bar{\beta}=\left\|\begin{bmatrix}
    \beta(1)\\
    \beta(0)
    \end{bmatrix} \right\|$.
\end{theorem}
\begin{proof}
See Appendix \ref{Ap:main} for the proof.
\end{proof}

From Theorem \ref{T:main} it follows that the deviation between the desired consensus value $x_\infty=\textbf{1}w_\ell\T \textbf{x}_0$ and the achieved consensus value $\tilde{x}_\infty$ decreases exponentially with the number of available bits $b$ and increases sub-linearly with the number of agents $N$.
The network topology among the agents (i.e., the $L$ matrix) affects the bound in \eqref{eq:consensus_deviation} through the quantity $\eta$ since $\eta > |\lambda_2|.$ 
Note that $b$ in \eqref{eq:consensus_deviation} depends on $\eta$ through \eqref{eq:b_eta}.

In \eqref{eq:consensus_deviation}, note that the right-hand-side is an increasing function of $\eta$ for $\eta \in (|\lambda_2|,1)$. 
Thus, to reduce the  deviation $\|\tilde{x}_\infty-x_\infty\|$, a lower value of $\eta $ is desired.
At the same time, from \eqref{eq:b_eta},  $b$ increases as $\eta$ is reduced.
This shows a desired trade-off between the quantization accuracy and the  deviation $\|\tilde{x}_\infty-x_\infty\|$.
While Theorem~\ref{T:main} proves that consensus is achieved under finite bit-rate constraint, we now state the following theorem which proves that the ICC algorithm under a bit-rate constraint is 
\ep{(1-\gamma)^2}.
\begin{theorem}[Main Result] \label{T:finall}
Under the assumption that the number of quantization bits $b$ satisfies \eqref{eq:b_eta} for some $\eta \in (|\lambda_2|,1) $, the ICC algorithm with finite bit-rate constraint is \ep{(1-\gamma)^2}.
\end{theorem}
\begin{proof}
See Appendix \ref{AP:T_finall} for the proof.
\end{proof}

We conclude this section with the following remark that presents the total bit-rate required for a network to achieve consensus.
\begin{rem}
Recall from Section \ref{sec:quant_prel} that for a network with $\N=\sum_{i\in \V}\N_i$  communication links and a total communication bit-rate of $\mathcal{B}_0$ bits per time-step, the available bits for each transmission is $\lfloor\mathcal{B}_0/\N \rfloor-b_0$, where $b_0$ is the number of overhead bits required for each transmission.
Therefore, from Lemma~\ref{L:rate_bound}, a sufficient condition for the network $\G$ to achieve consensus with  $(1-\gamma)^2$-\textit{protection} under bit-rate constraint of $\mathcal{B}_0$ bits per time-step is 
\begin{align}
\mathcal{B}_0> \N\left(b_0+\frac{1}{2}\log_2 N+\log_2{\left(\frac{7-3|\lambda_2|}{1-|\lambda_2|}\right)} \right). \label{eq:required_total_bit_rate}
\end{align}
\end{rem}
In summary, the analysis of this section shows that the following BICC (Bit-rate Constrained Innovation-Communication Consensus) algorithm (Algorithm~\ref{AL:BICC})  achieves \ep{(1-\gamma)^2} consensus as long as \eqref{eq:required_total_bit_rate} is satisfied (or equivalently, there exists $\eta \in (|\lambda_2|,1)$ such that $\mathcal{B}_0 \ge \N(b_0+b)$ where $b$ satisfies \eqref{eq:b_eta}).
\begin{algorithm}  
\caption{Bit-rate Constrained ICC (BICC)} \label{AL:BICC}
\begin{algorithmic}[1]
\STATE Initialize {$\small {\xi_i(-1)}=x_{i0}$} and $x_i(0)=x_{i0}^q$ for all $i\in \V$\\
\FOR{$k \in \bbN_0$}
\FORALL{$i=1,2,\ldots, N$}
\STATE agent $i$ sends ${\xi_i(k-1)}$ to $j\in \N^i$, 
\STATE agent $i$ receives ${\xi_j^q(k-1)}$ to $j\in \N_i$,
\FORALL{$j\in \N_i$}
\STATE $\hat x_j^i(k)=\hat x_j^i(k-1)+{\xi_j^q(k-1)},\quad \hat x_j^i(-1)=0$,
\ENDFOR
\STATE ${\xi_i(k)}=\sum_{j\in  \N_i}\kappa_{ij}(\hat x_j^i(k)-x_i(k))$
\STATE $x_i(k+1)=x_i(k)+{\xi_i^q(k)}$
\ENDFOR
\ENDFOR
\end{algorithmic}
\end{algorithm}

\section{Simulation Results} \label{Sec:simulation}
\begin{figure}
\centering
     \begin{tikzpicture}[scale=1]
     \centering
    \foreach \x in {0,1,...,25} {
    \node[circle,fill=red,inner sep=0 pt,minimum size=5pt]  (\x) at (\x*14.4:2.9) {};
        }
    \foreach \x in {0,...,23}
    {
    \pgfmathsetmacro{\y}{\x+2}
    \draw[->,>=stealth',gray] (\y) to [out=180+\x*14.4,in=270+\x*14.4] (\x);
    }
    \draw[->,>=stealth',gray] (1) to [out=180+24*14.4,in=270+24*14.4] (24);
    \end{tikzpicture}
    \caption{A network of $25$ agents where $\N_i=\{j: (i-j)\!\!\mod 25 = 2\}$.} \label{fig:small_network_schematic}
\end{figure}
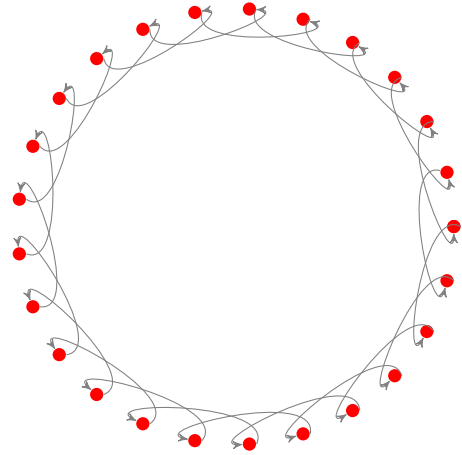

We simulate the ICC algorithm on a network $\G$ of $25$ agents. 
For this numerical experiment, the initial states of the agents are chosen randomly within the interval $[4,6]$. 
The connectivity $\Ed$ of this network forms a ring structure where agent $i$ sends information to agent $i+2$ and receives information from agent $i-2$, i.e., agent $1$ receives information from agent $24$ and sends information to agent $3$ and so forth.
A schematic diagram of the network is shown in Fig. \ref{fig:small_network_schematic}.
The weights  $\{\kappa_{ij}\}_{i,j \in \V}$ of the network are generated randomly, and the adversarial eavesdropping probability $\gamma$ is chosen to be $0.5$.

\begin{figure}
    \includegraphics[width=1 \linewidth]{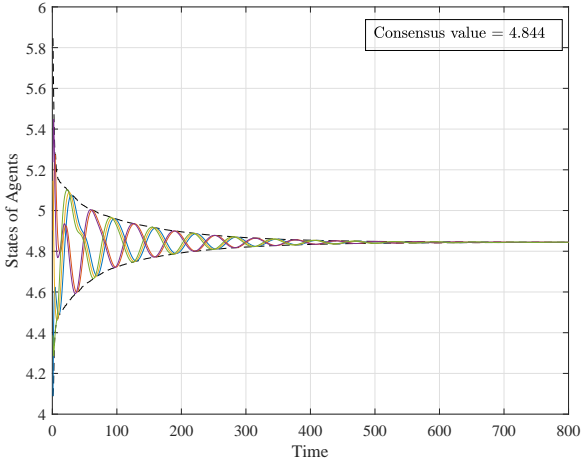}
    \caption{State trajectories $x_i(k)$ of $5$ agents are shown using solid colored lines. 
    The upper and lower black dashed lines represent the quantities $\max_i x_i(k)$ and $\min_i x_i(k)$, respectively.
    The final consensus value is 4.844.}
    \label{fig:trajectory_small_network}
\end{figure}

\begin{figure}
    \centering
    \includegraphics[width=\linewidth]{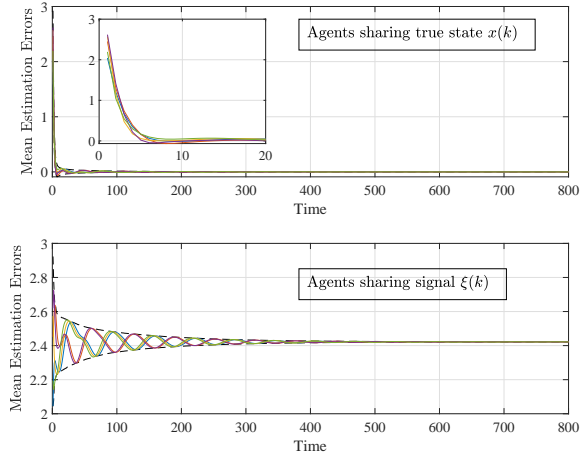}
    \caption{Top: Mean estimation error $\E[\ead_i(k)]$ when agents following protocol \eqref{eq:consensus} share their true state value. We zoom within the time interval $[0,20]$ for a better visualization.
    Bottom: Mean estimation error $\E[\ead_i(k)]$ when agent $i$ following Algorithm~\ref{AL:ICC} shares $\xi_i(k)$.} \label{fig:small_network_mean_error}
\end{figure}

\begin{figure}
    \centering
    \includegraphics[width=\linewidth]{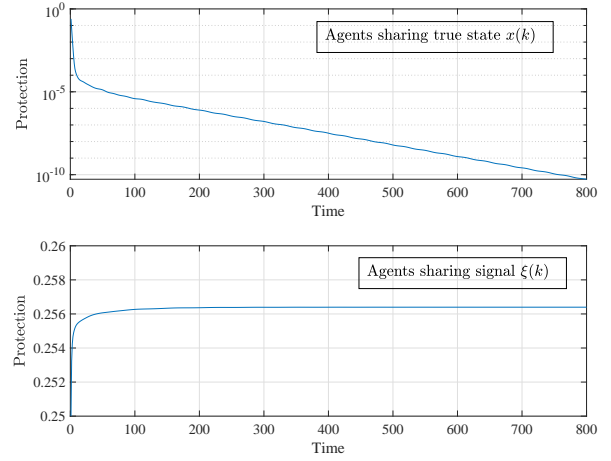}
    \caption{Top: $\E[\ead(k)\T\ead(k)]$ is plotted over time. Since $\E[\ead(k)\T\ead(k)]$ converges to zero fast, we use a logarithmic scale to better demonstrate its trajectory. The agents use consensus protocol \eqref{eq:consensus} and share their  state. Bottom: $\E[\ead(k)\T\ead(k)]$ is plotted over time for the case when the agents use the ICC protocol and share the signal $\xi(k)$.} \label{fig:small_network_quadratic_error}
\end{figure}

State trajectories for $5$ randomly selected agents are plotted in Fig.~\ref{fig:trajectory_small_network}.
In the same figure, we also plot the quantities $\max_ix_i(k)$ and $\min_i x_i(k)$ using black dashed lines. 
These black dashed lines represent the \textit{envelope} of the agents' states, i.e., the state trajectory of any agent remains within these two lines.
In Fig.~\ref{fig:small_network_mean_error} we plot mean estimation errors $\E[\ead_i(k)]$ for $5$ randomly chosen adversaries.
In the same figure, we also plot the mean estimation error envelope, i.e.,  $\max_i \E[\ead_i(k)]$ and $\min_i \E[\ead_i(k)]$ using black dashed lines.
To compare the performances of the consensus protocol \eqref{eq:consensus} and the ICC protocol of Algorithm \ref{AL:ICC}, we use two subplots in Fig.~\ref{fig:small_network_mean_error} to plot the mean estimation errors under both these protocols.
As seen in Fig.~\ref{fig:small_network_mean_error}, the mean estimation errors go to zero when the agents follow the consensus protocol \eqref{eq:consensus}. 
In contrast, the mean estimation errors converge to the vector $(1-\gamma)x_\infty$ when the agents follow the ICC protocol.
We also plot the \textit{protection level} ($\E[\ead(k)\T\ead(k)]/\|x(k)^2\|$) from adversarial eavesdropping  in Fig.~\ref{fig:small_network_quadratic_error}. 
We observe that the ICC protocol is at least $(1-\gamma)^2$-\textit{protected} (since $\gamma =0.5$) while the consensus protocol \eqref{eq:consensus} is $0$-\textit{protected}.

\begin{figure}
    \centering
    \includegraphics[width=\linewidth]{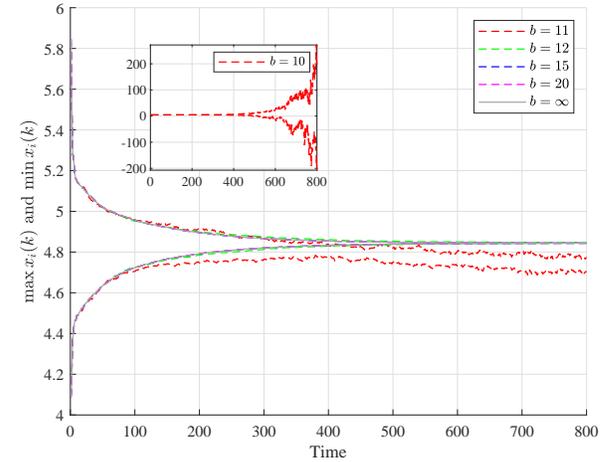}
    \caption{The upper and lower dashed lines  represent $\max_i x_i(k)$ and $\min_i x_i(k)$, respectively. The interval $[500,800]$ is zoomed in for better visualization. 
    Different colors represent different values of $b$.}
    \label{fig:quantization_b}
\end{figure}

Figs.~\ref{fig:trajectory_small_network}--\ref{fig:small_network_quadratic_error} show the performance of the system without any communication constraint.
Next, we study the same network while considering a finite bit-rate constraint.  
For this problem, we obtain $|\lambda_2|=0.992$ and hence, the right-hand-side of \eqref{eq:data_rate} evaluates to $11.2974$. 
Therefore, $b\ge 12$ is a sufficient condition to ensure consensus according to Lemma \ref{L:rate_bound}. 
We conducted several experiments by choosing the values of $b$ to be $10,11,12,15$ and $20$. 
The results in Figs.~\ref{fig:trajectory_small_network}--\ref{fig:small_network_quadratic_error} correspond of the case without any quantization, i.e.,  $b=\infty$. 
In Fig.~\ref{fig:quantization_b} we plot the consensus envelope ($\max_i x_i(k)$ and $\min_i x_i(k)$) for different values of $b$.
We observe that for $b=12,15$ and $20$, the performance is very similar to the unquantized case. 
The consensus values for $b=12,15$ and $20$ are $4.8455, 4.8436$ and $4.8439$, respectively. 
Given that consensus value is $4.844$ for the unquantized case, we notice that the effects of the finite bit-rate is practically indistinguishable.
We notice that for $b=11$, the consensus envelope deviates  from the true consensus envelope ($b=\infty$).
Furthermore, in this case, consensus is not achieved as the consensus envelope does not converge at a single value.
We also considered a case when $b=10$, and we see that the consensus dynamics becomes unstable and the consensus envelope diverges. 
This is plotted in the subplot within Fig.~\ref{fig:quantization_b}.

\begin{figure}
    \centering
    \includegraphics[width=\linewidth]{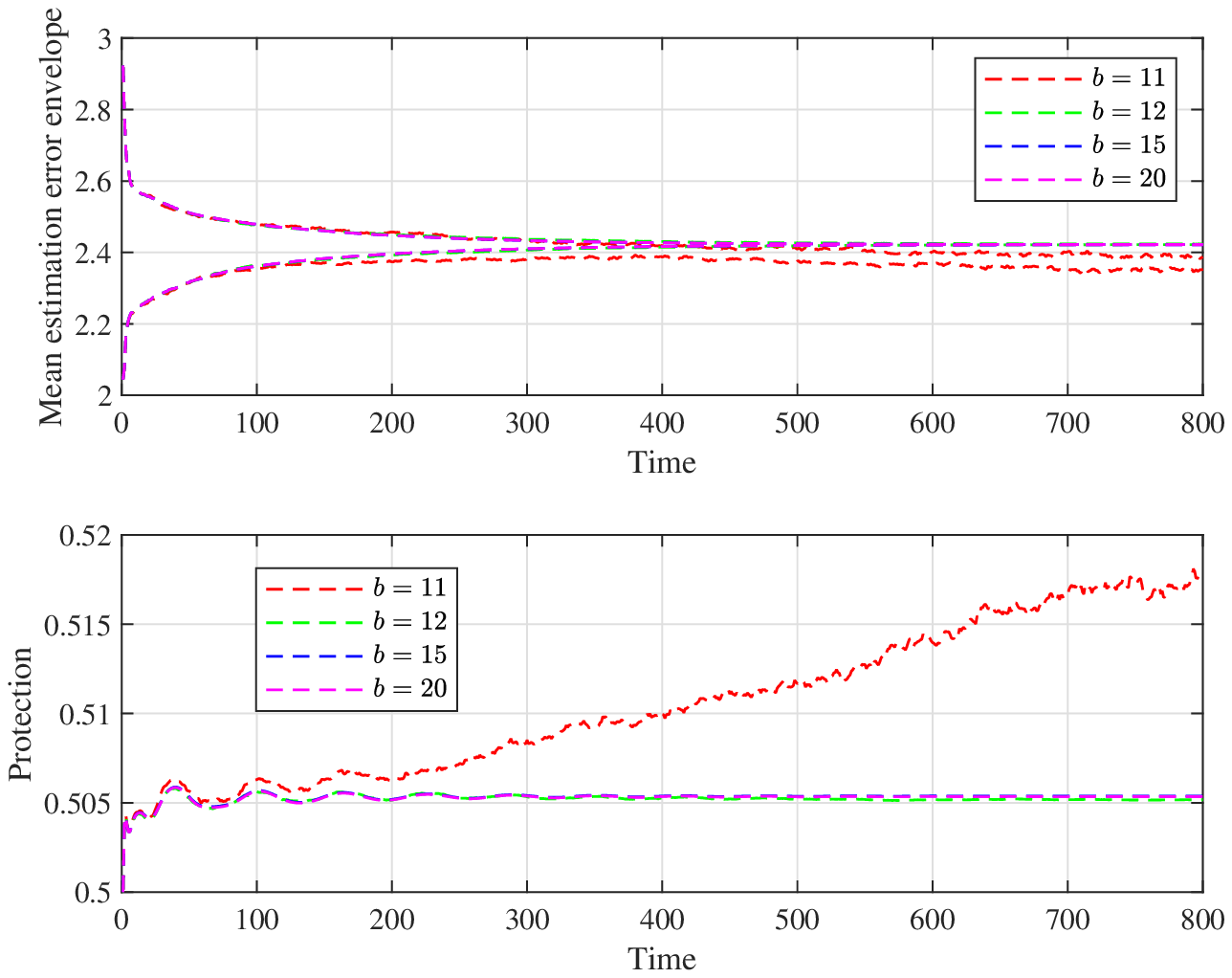}
    \caption{Top: Mean estimation error envelop $\max_i\E[\ead_i(k)]$ and $\min_i\E[\ead_i(k)]$ for different values of $b$.
    Bottom: The trajectory of $\E[\ead(k)\T \ead(k)]$ is plotted for different values of $b$.}
    \label{fig:sigma_with_quantization}
\end{figure}

In Fig.~\ref{fig:sigma_with_quantization}, we plot the mean estimation error envelope ($\max_i\E[\ead_i(k)]$ and $\min_i\E[\ead_i(k)]$) and the protection amount $\E[\ead(k)\T\ead(k)]/\|x(k)\|^2$ for $b=11,12,15$ and $25$ and we notice that for every choice of $b$, the proposed ICC algorithm ensures at least $(1-\gamma)^2 = 0.25$ protection.
In fact, the simulated amount of protection is almost twice of the theoretically predicted amount.
Interestingly, for $b=11$, the amount of protection is higher than any other case. However, in this case the agents do not achieve consensus.

\section{Conclusion} \label{Sec:Conclusion}
In this work we revisit the consensus problem among a group of agents.
We consider an adversarial model where the adversaries are not participating in the consensus, but rather, they are monitoring and intercept intermittently the inter-agent communications. 
The adversarial eavesdropping mechanism is assumed to be imperfect and the agents exploit this imperfection to devise a new consensus protocol.
We define a metric for protection against such adversaries and present an \ep{\epsilon} consensus algorithm that does not require the agents to share their true state values in order to achieve consensus. 
We show that the proposed consensus algorithm  does not affect the consensus value.
Existing consensus protocols to achieve privacy lack formal guarantees of consensus with finite data-rate constraint.
We explicitly consider the bit-rate constraint of the whole network and design dedicated dynamic quantizers for the agents to ensure that consensus can be achieved if the bit-rate is higher than a lower bound that depends on the number of agents and the graph matrix $L$.

\section{Acknowledgements}

The authors greatly appreciate the valuable comments and suggestions for improvement from Anastasios Tsiamis and George J. Pappas on an earlier version of this manuscript. 


\appendices \label{Appendix}
\section{Proof of Theorem \ref{T:wl}}\label{AP:wl}
Let us consider the Jordan decomposition of the matrix $L$
\begin{align} \label{eq:ap_prejordan}
    L=Q\J R,
\end{align}
where $Q=R^{-1}$ and $\J$ is the Jordan matrix. 
Since the communication graph $\G$ is strongly connected, the unit eigenvalue of $L$ has algebraic and geometric multiplicity  one \cite{olfati2004consensus,xiao2007distributed}. 
Thus, $\J$ can be written as
\begin{align}
    \J=\begin{bmatrix}1 & 0_{1\times N-1}\\
    0_{N-1\times 1} & \J_1
    \end{bmatrix}, \nonumber
\end{align}
for some matrix $\Lambda\in \R^{n-1\times n-1}$.
Thus, we can write $L$ as
\begin{align}
    L&=\begin{bmatrix}q_1,q_2,\ldots,q_N
    \end{bmatrix}\begin{bmatrix}1 & 0_{1\times N-1}\\
    0_{N-1\times 1} & \J_1
    \end{bmatrix}
    \begin{bmatrix} 
    r_1\T\\
    r_2\T\\
    \vdots\\
    r_N\T
    \end{bmatrix}\nonumber \\
    &=q_1r_1\T +\begin{bmatrix}q_2,\ldots,q_N
    \end{bmatrix}\J_1 
    \begin{bmatrix} 
    r_2\T\\
    \vdots\\
    r_N\T
    \end{bmatrix}\nonumber \\
    &\triangleq q_1r_1\T+L_1, \label{Eq:AP_jordan}
\end{align}
where $r_i\T$ is the $i$-th row of $R$ and $q_j$ is the $j$-th column of $Q$ and where we have used the fact that $Q=R^{-1}$ implies $r_i\T q_j= 1$ if $i=j$ and $r_i\T q_j=0$ if otherwise.
Since $L$, $Q$ and $R$ are related by the Jordan decomposition \eqref{eq:ap_prejordan}, each $q_i$ is a right eigenvector of $L$ and each $r_i\T$ is a left eigenvector of $L$.
Furthermore, since $r_i\T q_1=0$ for all $i\ne 1$, we obtain from \eqref{Eq:AP_jordan} that 
\begin{align*}
    L q_1=q_1,
\end{align*}
implying that $q_1=\textbf{1}q$ for some $q\in \R$.
By defining $w_\ell=qr_1$, we obtain
$L=\textbf{1}w_\ell\T+L_1$ where $w_\ell\T L_1=0$ and $w_\ell\T\textbf{1}\!=\!r_1\T q_1\!=\!1$. \\
\qed
\section{Proof of Proposition \ref{Pro:prel3}}\label{AP:Pro_prel3}
Statements (i) and (ii) follow from Theorem~\ref{T:wl}.
We prove statement (iii)  by induction.
Clearly, (iii) holds for $k=1$. 
Now, assume that it holds for some $k>1$. 
Then, by using (i) and (ii),
\begin{align*}
    (L-J)^{k+1}&=(L^k-J^k)(L-J)\\
    &=L^{k+1}-J^kL-L^kJ+J^{k+1}=L^{k+1}-J^{k+1}.
\end{align*}

For statement (iv), we observe that $(L-I)J^k=LJ^k-J^k=J-J=0$ from (i) and (ii). 
Consequently, by using (iii), we  obtain
\begin{align*}
    (L-I)L^k&=(L-I)L^k-(L-I)J^k=(L-I)(L^k-J^k)\\
    &=(L-I)(L-J)^k.
\end{align*}

For statement (v),  notice from the proof of Theorem~\ref{T:wl} that
$L=J+L_1$, where the eigenvalues of $L_1$ are in the set $\{\lambda_2,\ldots,\lambda_N\}$.
Due to Theorem \ref{T:prel1}, we have that $|\lambda_2| =\max\{|\lambda_2|,\ldots,|\lambda_N|\}$ and hence the spectral radius of $J_1=L-J$ satisfies $\rho(L-J)=\rho(L_1)=|\lambda_2|$.

 Finally, (vi) is a direct consequence of (v). \qed

\section{Proof of Theorem \ref{T:0-protection}}\label{AP:0-protection}
Let $M(k)\triangleq I- \diag{\mu_1(k),\cdots,\mu_N(k)}$. 
Then,  $\xad(k)$ and the estimation error $\ead(k)=x(k)-\xad(k)$ follow the dynamics
\begin{align}
    \xad(k)&=(I-M(k))x(k)+M(k)\xad(k-1), \nonumber\\
    \ead(k)&=M(k)\big(\ead(k-1)+x(k)-x(k-1)\big)\nonumber\\
   &= M(k)\ead(k-1)+M(k)(L-I)x(k-1) \label{eq:ead}.
\end{align}

In order to prove the claim that the classical communication scheme (that transmits the states of the agents) leads to  $0$-\textit{protection}, we first show that the random variable $\ead(k)$ converges to $0$ in the mean-square sense as $k$ goes to infinity.
Taking expectation on both sides of \eqref{eq:ead}, we obtain
\begin{align}
    \E[\ead(k)]=(1-\gamma)\E[\ead(k-1)]+(1-\gamma)(L-I)x(k-1),
\end{align}
where we have used the fact\footnote{
Since $\ead(k-1)$ is a function of $\{M(1),\ldots,M(k-1)\}$ and $M(k)$ is independent of the past $\{M(s)\}_{s< k }$, we have that $\ead(k-1)$ and $M(k)$ are uncorrelated, i.e., $\E[M(k)\ead(k-1)]=\E[M(k)]\E[\ead(k-1)]$.} that $M(k)$ is uncorrelated to $\ead(k-1)$.
Since $0<\gamma<1$ and $\lim_{k\to \infty} x(k)\overset{}{\to} x_\infty$,  $\E[\ead(k)]$ converges to a finite value. 
In fact,
\begin{align*}
    \lim_{k\to \infty} \E[\ead(k)]=&(1-\gamma) \lim_{k\to \infty}\E[\ead(k-1)]+\\
    &(1-\gamma)(L-I) \lim_{k\to \infty}x(k-1)\\
    \overset{(a)}{=}&(1-\gamma)\lim_{k\to \infty}\E[\ead(k-1)],
\end{align*}
where $(a)$ follows from the fact that $x_\infty=Lx_\infty$, and hence
\begin{align*}
\lim_{k\to \infty} \E[\ead(k)]=&0.
\end{align*}
Furthermore, we have that
\begin{align}
    \Sigma^{\rm ad}(k)&\triangleq \E[\ead(k){\ead}\T(k)] \nonumber \\
    &=(1-\gamma)^2\sad(k-1)\nonumber\\
    &~~+(1-\gamma)^2(L-I)x(k-1)x\T(k-1)(L\T-I) \nonumber \\
    &~~+(1-\gamma)^2(L-I)x(k-1)\E[\ead(k-1)]\T \nonumber\\
    &~~+(1-\gamma)^2\E[\ead(k-1)]x\T(k-1)(L\T-I) \nonumber \\
    &~~+\gamma(1-\gamma)\diag{\{(x_i(k)-x_i(k-1))^2\}_{i\in \V}}, \label{eq:sigma_ad}
\end{align}
where in deriving \eqref{eq:sigma_ad} we have used the fact that the $ij$-th element of the matrix $M(k)(L-I)x(k-1)x(k-1)\T (L-I)\T M(k)\T$ is $(1-\mu_i(k))(1-\mu_j(k))(x_i(k)-x_i(k-1))(x_j(k)-x_j(k-1))$, and its expected value is $(1-\gamma)^2(x_i(k)-x_i(k-1))(x_j(k)-x_j(k-1))$ if $i\ne j$ and $(1-\gamma)(x_i(k)-x_i(k-1))^2$ if otherwise.

Since $(1-\gamma)^2<1$, $\Sigma^{\rm ad}(k)$ has  stable dynamics and as the last four terms in \eqref{eq:sigma_ad} converge to the zero matrix, it follows that $\Sigma^{\rm ad}(k)$ also converges.
Furthermore, by observing that $\lim_{k\to \infty} (L-I)x(k)=0$, we have from \eqref{eq:sigma_ad} that $\lim_{k\to \infty} \Sigma^{\rm ad}(k) = (1-\gamma)^2 \lim_{k\to \infty} \Sigma^{\rm ad}(k-1)=0$.
Since $\E[\ead(k)\T\ead(k)]=\tr(\sad(k))$ where, $\tr(\cdot)$ denotes the trace of a matrix, we obtain 
\begin{align*}
    \lim_{k\to \infty}\E[\ead(k)\T\ead(k)]=0,
\end{align*}
and consequently, the consensus protocol \eqref{eq:consensus} is (\textit{asymptotically}) $0$-\textit{protected}.  \qed

\section{Proof of Theorem \ref{T:ICC2}} \label{Ap:ICC2}
The group of agents $\{i_1,\cdots,i_\ell\}=\N^j$ receive measurements from agent $j$.
The ICC algorithm (line 6, Algorithm~\ref{AL:ICC}) requires that each agent $i_m \in \N^j$   computes a local estimate ($\hat x_j^{i_m}(k)$) of the state $x_j(k)$ by following the dynamics
\begin{align*}
    &\hat x_j^{i_m}(k+1)=\hat x_j^{i_m}(k)+\xi_j(k),\\
    &\hat x_j^{i_m}(-1)=0.
\end{align*}
Based on the above dynamics, it is easy to see that $\hat x_j^{i_1}(k)=\cdots=\hat x_j^{i_\ell}(k)$.
Therefore, the estimates of the state $x_j(k)$ performed independently by the agents $\{i_1,\cdots,i_\ell\}$ actually coincide, and hence, we will simply write $\hat x_j(k)$ to denote the estimate of $x_j(k)$ performed by any agent $i$ such that $i\in \N^j$. 

Next, we show that, for all $k\in \bbN_0$ and $i\in \V$, $\hat x_i(k)=x_i(k)$.
First, notice that
\begin{align}
    \hat x_j(k+1)=&\hat{x}_j(k)+\xi_j(k) , \label{eq:hat_x_j_dynamics}\\
    \hat{x}_j(0)=\xi_j(&-1)=x_j(0), \nonumber
\end{align}
and from Algorithm~\ref{AL:ICC}, the dynamics of agent $j$ is
\begin{align} \label{eq:x_j_dynamics}
    x_j(k+1)=x_j(k)+{\xi_j(k)}.
\end{align}
By comparing equations \eqref{eq:hat_x_j_dynamics} and \eqref{eq:x_j_dynamics}  we have that $\hat  x_j(k)=x_j(k)$ for all $k\in \bbN_0$ and $j\in \V$.
Therefore, it follows that $\xi_i(k)=\sum_{j\in \N_i}\kappa_{ij} (\hat x_j(k)-x_i(k))=\sum_{j\in \N_i}\kappa_{ij}( x_j(k) -x_i(k))$, and hence the ICC algorithm update step  $x_i(k+1)=x_i(k)+{\xi_i(k)}$ (line 9, Algorithm~\ref{AL:ICC}) is equivalent to $$x_i(k+1)=x_i(k)+\sum_{j\in \N_i}\kappa_{ij} (x_j(k)-x_i(k)).$$
 Therefore, the evolution of the state $x(k)$ executing the ICC algorithm is the same as the state following the dynamics \eqref{eq:discrete_consensus},  and hence, $\lim_{k\to \infty} x(k)\!=\!x_\infty$.
This completes the proof.\qed

\section{Proof of Theorem \ref{T:ICC}} \label{Ap:ICC}
The dynamics of agent $i$ executing Algorithm~\ref{AL:ICC} is 
\begin{align} \label{eq:AP_dyn}
    x_i(k+1)=x_i(k)+\xi_i(k),
\end{align}
where $\xi_i(k)$ is the new information, or \textit{innovation}, that agent $i$ accumulates by combining the incoming data from its neighbors $\N_i$, and then broadcasting it to its outgoing neighbors through the links $E_{ij}$. 
Let $\mathcal I_i(k)=\{y_i(0),y_i(1),\ldots,y_i(k) \}$ denote the information available to adversary $i$ at time $k$, where
\begin{align} 
    y_i(k)=\begin{cases}
    \xi_i(k-1), & \text{if } \mu_i(k)=1,\\
    \varnothing, & \text{if } \mu_i(k)=0,
    \end{cases} \nonumber  
\end{align}
where $\varnothing$ denotes absence of a measurement and $\mu_i(k)$ is defined in \eqref{eq:mu_bernoulli}.
Recall that the adversaries try to minimize  $ \E[\ead(k)\T\ead(k)]$, which quantifies the amount of protection according to Definition~\ref{Df:e-protected}.
If $\xad_i(k)$ is the estimate of $x_i(k)$ performed by adversary $i$ based on the information $\mathcal I_i(k)$ to minimize $\E[\ead_i(k)^2]$, then the optimal estimate is $\xad_i(k)=\E[x_i(k)|\mathcal{I}_i(k)]$. 
Therefore, the estimation dynamics for the adversary becomes
\begin{align}\label{eq:ap_adv_est_dyn2}
    \xad_i(k+1)=&\xad_i(k)+\mu_i(k)\xi_i(k).
\end{align}
Consequently, the estimation error $\ead(k)$ has the dynamics
\begin{align}
    \ead(k+1)=\ead(k)+M(k)\xi(k), \nonumber \quad \ead(0) = M(0)\xi(-1),
\end{align}
where $M(k)\!\triangleq \!I\!-\!\diag{\mu_1(k),\cdots,\mu_N(k)}$.
The mean $\E[\ead(k)]$ and the second moment $\sad(k)\triangleq  \E[(\ead(k)\ead(k)\T]$ of the error follow the dynamics
\begin{align}
    &\E[\ead(k+1)]=\E[\ead(k)]+(1-\gamma)\xi(k) \label{eq:AP_e_dynamics} \\
    & \E[\ead(0)] =(1-\gamma)x(0) \nonumber \\
    &\sad(k+1)=\sad(k) +\Gamma(k)\label{eq:AP_sigma}\\
    &\sad(0) = (1-\gamma)^2x(0) x(0)\T \\
    &~~~~~~~~+\gamma(1-\gamma)\diag{\xi_1(0)^2,\ldots,\xi_N(0)^2} \nonumber \\
    &\Gamma(k)= (1-\gamma)^2\xi(k)\xi(k)\T \nonumber \\
    &~~~~\quad+(1-\gamma)\left(\E[\ead(k)]\xi(k)\T+\xi(k)\E[\ead(k)\T] \right) \nonumber\\
    &~~~~\quad+\gamma(1-\gamma)\diag{\xi_1(k)^2,\ldots,\xi_N(k)^2}, \label{eq:AP_gamma_k}
\end{align}
where we have used the fact that the $ij$-th element of the matrix $M(k)\xi(k)\xi(k)\T M(k)\T$ is $(1-\mu_i(k))(1-\mu_j(k))\xi_i(k)\xi_j(k)$ and 
\begin{align*}
    \E[(1-\mu_i(k))(1-\mu_j(k))&\xi_i(k)\xi_j(k)] \\
    &= \begin{cases}
    (1-\gamma)^2\xi_i(k)\xi_j(k),  &\text{ if } i\ne j,\\
    (1-\gamma)\xi_i(k)^2,  &\text{ if } i=j.
    \end{cases}
\end{align*}
Notice from \eqref{eq:AP_dyn} that 
$x(k+1)=x(k)+\xi(k)=Lx(k)$ and thus
\begin{align}
   \xi(k)=(L-I)x(k)=(L-I)L^k x(0) \label{eq:AP_xi_x_k}.
\end{align}
From the dynamics of $\E[\ead(k)]$ given in \eqref{eq:AP_e_dynamics}, we obtain 
\begin{align}
    \E[\ead(k+1)]&= \E[\ead(0)] + (1-\gamma)\sum_{t=0}^{k}\xi(t) \nonumber \\ \nonumber
    &=(1-\gamma)x(0)+ (1-\gamma)(L-I) \sum_{t=0}^k L^t x(0)\\
    &=(1-\gamma) L^{k+1} x(0). \label{eq:AP_e_x_k}
\end{align}
Using \eqref{eq:AP_xi_x_k}  and \eqref{eq:AP_e_x_k}, we may simplify \eqref{eq:AP_gamma_k} as follows
\begin{align*}
    \Gamma(k) =& (1-\gamma)^2((L-I)L^k x(0))((L-I)L^k x(0))\T \nonumber \\
    &+(1-\gamma)^2(L^k x(0))((L-I)L^k x(0))\T\\
    &+(1-\gamma)^2((L-I)L^k x(0))(L^k x(0))\T \\
    &+\gamma(1-\gamma)\diag{\xi_1(k)^2,\ldots,\xi_N(k)^2} \\
    =&(1-\gamma)^2(L^{k+1}x(0)) (L^{k+1}x(0))\T  \\
    &-(1-\gamma)^2(L^{k}x(0)) (L^{k}x(0))\T \\
    &+\gamma(1-\gamma)\diag{\xi_1(k)^2,\ldots,\xi_N(k)^2}\\
    =& (1-\gamma)^2\big(x(k+1)x(k+1)\T -x(k) x(k)\T \big)\\
    &+\gamma(1-\gamma)\diag{\xi_1(k)^2,\ldots,\xi_N(k)^2}.
\end{align*}
Therefore, from \eqref{eq:AP_sigma}, we obtain
\begin{align*}
    &\sad(k)=\sad(0) + \sum_{t=0}^{k-1}\Gamma(t)\\
    &= (1-\gamma)^2 x(k) x(k)\T + \gamma(1-\gamma)\sum_{t=0}^{k-1} \diag{\xi_1(t)^2,\ldots,\xi_N(t)^2}.
\end{align*}
Thus, for all $k$, $\E[\ead(k)\T\ead(k)]=\tr(\sad(k)) \ge (1-\gamma)^2\|x(k)\|^2 +c $ where $c=\gamma(1-\gamma)\|x(0)\|^2$,  and hence,
the ICC algorithm is \ep{(1-\gamma)^2}. \qed

\section{Proof of Lemma \ref{L:rate_bound}} \label{AP:L_rate_bound}
From the expression of $\beta(k)$ in \eqref{eq:beta_dynamics} one observes that for all $k\in \bbN$,
\begin{align*}
    &\frac{\beta(k+1)}{2}=\frac{3\sqrt{N} \beta(k)}{2^{b+1}}
    +\frac{4\sqrt{N}}{2^{b+1}}\sum_{\ell=0}^{k-1}|\lambda_2|^{k-1-\ell}\beta(\ell) \\
    &\quad ~~~~~~~~~~~~+4|\lambda_2|^{k-1}\sqrt{N}\max\{|x_{\min}|,|x_{\max}|\},
\end{align*}
\begin{align*}
    \frac{\beta(k+1)}{2}&-\frac{3\sqrt{N} \beta(k)}{2^{b+1}}
    =\frac{4\sqrt{N}}{2^{b+1}}\sum_{\ell=0}^{k-1}|\lambda_2|^{k-1-\ell}\beta(\ell)\\
    &+4|\lambda_2|^{k-1}\sqrt{N}\max\{|x_{\min}|,|x_{\max}|\}\nonumber\\
    &=\frac{4\sqrt{N}}{2^{b+1}}\beta(k-1)+|\lambda_2|\left(\frac{\beta(k)}{2}-\frac{3\sqrt{N} \beta(k-1)}{2^{b+1}}\right).
\end{align*}
By rearranging terms, we obtain
\begin{align*}
    &\beta(k+1)=\left(\frac{3\sqrt{N}}{2^{b}}+|\lambda_2|\right)\beta(k)+\frac{\sqrt{N}}{2^{b}}(4-3|\lambda_2|)\beta(k-1),\\
    &\text{and thus,} \hfill\\
    &\begin{bmatrix}
    \beta(k+1)\\
    \beta(k)
    \end{bmatrix} =
     \begin{bmatrix}
    \frac{3\sqrt{N}}{2^{b}}+|\lambda_2| ~~& \frac{\sqrt{N}}{2^{b}}(4-3|\lambda_2|)\\
    1 & 0
    \end{bmatrix}
     \begin{bmatrix}
    \beta(k)\\
    \beta(k-1)
    \end{bmatrix}.
\end{align*}
Therefore, $\beta(k)$ follows a linear dynamics with initial conditions given in \eqref{eq:alpha_initial}-\eqref{eq:beta_initial}, and the dynamics is stable if and only if the eigenvalues of the transition matrix lie inside the unit circle.
Thus, $\beta(k)\to 0$ exponentially if
\begin{align*}
    \frac{3\sqrt{N}}{2^{b}}+|\lambda_2|+\sqrt{\left(\frac{3\sqrt{N}}{2^{b}}+|\lambda_2|\right)^2\!\!
    +4\frac{\sqrt{N}}{2^{b}}(4-3|\lambda_2|)} ~< 2.
\end{align*}
The above condition can be simplified to the  condition 
\begin{align}
    2^{b}>  \frac{\sqrt{N}(7-3|\lambda_2|)}{1-|\lambda_2|},
\end{align}
which is equivalent to equation \eqref{eq:data_rate} of the lemma. \qed

\section{Proof of Lemma \ref{L:beta_rate}} \label{Ap:L_beta_rate}
In the proof of Lemma~\ref{L:rate_bound}, we notice that $\beta(k)$ satisfies the dynamics
\begin{align*}
    \begin{bmatrix}
    \beta(k+1)\\
    \beta(k)
    \end{bmatrix} =
     \begin{bmatrix}
    \frac{3\sqrt{N}}{2^{b}}+|\lambda_2| ~~& \frac{\sqrt{N}}{2^{b}}(4-3|\lambda_2|)\\
    1 & 0
    \end{bmatrix}
     \begin{bmatrix}
    \beta(k)\\
    \beta(k-1)
    \end{bmatrix}.
\end{align*}
One may verify that the maximum eigenvalue of the matrix $\begin{bmatrix}
    \frac{3\sqrt{N}}{2^{b}}+|\lambda_2| ~~& \frac{\sqrt{N}}{2^{b}}(4-3|\lambda_2|)\vspace*{6 pt}\\
    1 & 0
    \end{bmatrix}$ is $$\frac{1}{2}\left(\frac{3\sqrt{N}}{2^{b}}\!+\!|\lambda_2|+\sqrt{\left(\frac{3\sqrt{N}}{2^{b}}+|\lambda_2|\right)^2\!\!+4\frac{\sqrt{N}}{2^{b}}(4-3|\lambda_2|)}\right).$$
Substituting the value of $b$ from \eqref{eq:b_eta} yields 
$$\frac{3\sqrt{N}}{2^{b}}\!+\!|\lambda_2|+\sqrt{\left(\frac{3\sqrt{N}}{2^{b}}+|\lambda_2|\right)^2\!\!+4\frac{\sqrt{N}}{2^{b}}(4-3|\lambda_2|)} = 2\eta.$$
Therefore, from the linear dynamics of $[\beta(k+1),\beta(k)]\T$, one obtains
\begin{align}
    \left\|\begin{bmatrix}
    \beta(k+1)\\
    \beta(k)
    \end{bmatrix} \right\| \le \eta \left\|\begin{bmatrix}
    \beta(k)\\
    \beta(k-1)
    \end{bmatrix} \right\|\le \eta^k\left\|\begin{bmatrix}
    \beta(1)\\
    \beta(0)
    \end{bmatrix} \right\|.\nonumber
\end{align}
Finally, we can write $\|\beta(k)\|\le \bar{\beta}\eta^k $ where $\bar{\beta}=\left\|\begin{bmatrix}
    \beta(1)\\
    \beta(0)
    \end{bmatrix} \right\|$.
    Since $\beta(k)\ge 0$ for all $k$ as per \eqref{eq:beta_dynamics}, we have $\beta(k)=\|\beta(k)\|\le \bar{\beta}\eta^k$. \qed


\section{Proof of Lemma \ref{L:no_saturation}} \label{Ap:L_xi}
We prove this lemma by induction. 
To this end, we first show that the lemma holds for $k=0$ and $1$.
Each agent initializes $\xi_i(-1)=x_{i0}$ in \eqref{eq:xi_quant_2}.
At time $k=0$, $\xi_i(-1)$ is quantized using the quantizer $q_i(\cdot,0)$ with parameters $\alpha_i(0)=x_{\min}$ and $\beta(0)=(x_{\max}-x_{\min})$. 
Since $x_{\max}\ge x_{i0}\ge x_{\min}$ for all $i$, it follows that $\xi_i(-1)\in [\alpha_i(0),\alpha_i(0)+\beta(0)]$. 
Hence, the lemma is true for $k=0$.
At $k=1$, $\xi_i(0)=\sum_{j \in \N_i}\kappa_{ij}(x_j(k)-x_i(k))$ is quantized using the quantizer $q_i(\cdot,1)$.
Therefore, we obtain
\begin{align*}
    \xi_i(0)- \xi_i^q(-1) =& \sum_{j \in \N_i}\kappa_{ij}(x_j(0)-x_i(0)) - \xi_i^q(-1) \\
    =&  \sum_{j \in \N_i}\kappa_{ij}(x_{j0}^q-x_{i0}^q) - x_{i0}^q ,
\end{align*}
which further yields
\begin{align*}
    \|\xi_i(0)- \xi_i^q(-1)\| & \le (1+\sum_{j\in \N_i} \kappa_{ij})|x^q_{i0}| + \max_{j\in \N_i} |x^q_{j0}|\sum_{j\in \N_i} \kappa_{ij} \\
    & \le 3 \max\{|x_{\min}|, |x_{\max}|\},
\end{align*}
where we have used the fact that $\sum_{j\in \N_i} \kappa_{ij} \le 1$ for all $i$. 
From the expressions of $\alpha_i(1)$ and $\beta(1)$, we notice that $\xi_i(0) \in [\alpha_i(1),\alpha_i(1)+\beta(1)]$, and thus, the lemma holds for $k=1$.

Let us now assume that the lemma holds for all time-instances $0,1,\ldots,k_0$, i.e, $\xi_i(\ell-1) \in [\alpha_i(\ell),\alpha_i(\ell)+\beta(\ell)]$ for all $\ell=0,1,\ldots,k_0$.
Consequently, we have from \eqref{eq:delta_ik} that $|\Delta^q_i(\xi_i(\ell-1),\ell)|\le \beta(\ell)/{2^{b+1}}$ for all $\ell=0,1,\ldots,k_0$.
We will show that the lemma holds for time $k_0+1$.
In order to proceed,  note from \eqref{eq:xi_quant_2} that, for all $k\in \bbN_0$, 
\begin{align}
    \xi(k) = (L-I) x(k), \label{eq:xi_x_apx}
\end{align}
where $\xi(k)=[\xi_1(k),\ldots,\xi_N(k)]\T$.
For any $\xi \in \R^N$, let us define the vector $\Delta^q(\xi,k)=[\Delta^q_1(\xi_1,k), \ldots, \Delta^q_N(\xi_N,k)]\T$, where $\Delta^q_i(\cdot,k)=(\cdot)-q_i(\cdot,k)$ is the quantization error function of the dynamic quantizer $q_i(\cdot,k)$.
Thus, we obtain from \eqref{eq:dyn_quant_2} that $x_i(k+1)=x_i(k)+\xi_i(k)-\Delta_i^q(\xi_i(k),k+1)$, and hence, 
\begin{align}
    x(k+1) &= L x(k) - \Delta^q(\xi(k),k+1), \nonumber \\
    &=L^{k+1} x(0)-\sum_{\ell=0}^{k}L^{k-\ell}\Delta^q(\xi(\ell),\ell+1).  \label{eq:x_dynamics}
\end{align}
From \eqref{eq:xi_x_apx} and \eqref{eq:x_dynamics}, we obtain
\begin{align} \label{eq:xi_difference}
    \xi(k+1) &- \xi(k) = (L-I) (x(k+1)-x(k)) \nonumber \\
   & =(L-I)^2x(k) -(L-I)\Delta^q(\xi(k),k+1) \nonumber \\
   &=(L-I)^2\Big(L^{k}x(0)-\sum_{\ell=0}^{k-1}L^{k-1-\ell}\Delta^q(\xi(\ell),\ell+1) \Big) \nonumber \\
   &~~-(L-I)\Delta^q(\xi(k),k+1).
\end{align}
We therefore obtain that
\begin{align} \label{eq:xi_difference_quant}
    \|\xi&(k_0)- \xi^q(k_0-1)\| \nonumber \\ \nonumber
    &= \|(\xi(k_0)-\xi(k_0-1))+\Delta^q(\xi(k_0-1),k_0)\|\\ \nonumber
    &\overset{(a)}{ \le} \|(L-I)^2L^{k_0-1} x(0)\| \\ \nonumber
    &~~~~+\sum_{\ell=0}^{k_0-2}\|(L-I)^2 L^{k_0-2-\ell}\Delta^q(\xi(\ell),\ell+1)\|\\ \nonumber
    &~~~~+\|(2I-L)\Delta^q(\xi(k_0-1),k_0)\|\\ \nonumber
    &\overset{(b)}{\le} 4|\lambda_2|^{k_0-1}\|x(0)\|+ \frac{4\sqrt{N}}{2^{b+1}}\sum_{\ell=0}^{k_0-2}|\lambda_2|^{k_0-2-\ell}\beta(\ell+1) \\
    &~~~~+\frac{3\sqrt{N}\beta(k_0)}{2^{b+1}}
\end{align}
where, we have used \eqref{eq:xi_difference} to substitute $\|(\xi(k_0)-\xi(k_0-1))\|$ in deriving inequality $(a)$, and in deriving inequality $(b)$, we have first used Proposition~\ref{Pro:prel3} to replace $(L-I)^2L^t$ with $(L-I)^2(L-J)^t$ and further use that fact that $\rho(L-J)\le |\lambda_2|$. 
Then, using Theorem~\ref{T:prel1}, we observe that $\|L-I\| \le 2$ since all the eigenvalues of $L$ have magnitudes less than or equal to $1$, and similarly $\|2I-L\| \le 3$.
Finally, we have used the assumption that the lemma holds for all $\ell=0,\ldots,k_0$ to conclude that
$\|\Delta^q(\xi(\ell-1),\ell)\|\le \sqrt{N}\max_{i}|\Delta^q_i(\xi_i(\ell-1),\ell)| \le \sqrt{N}\beta(\ell)/2^{b+1}$.
Since $x_i(0)= x^q_{i0} = x_{i0} -\Delta^q_i(x_{i0}, 0)$, we obtain
\begin{align} \label{eq:x0_beta}
    \|x(0)\| \le &\sqrt{N}\max_i\Big(|x_{i0}| + |\Delta^q_i(x_{i0}, 0)| \Big) \nonumber \\
    \le & \sqrt{N} \max\{|x_{\min}|, |x_{\max}|\} +\frac{\sqrt{N}\beta(0)}{2^{b+1}}.
\end{align}
Combining \eqref{eq:xi_difference_quant} and \eqref{eq:x0_beta}, we obtain
\begin{align}  \label{eq:xi_final}
    \|\xi&(k_0)- \xi^q(k_0-1)\| \nonumber \\ \nonumber
    &\le 4\sqrt{N} \max\{|x_{\min}|, |x_{\max}|\} |\lambda_2|^{k_0-1} \\
    &~~~~+ \frac{4\sqrt{N}}{2^{b+1}}\sum_{\ell=0}^{k_0-1}|\lambda_2|^{k_0-1-\ell}\beta(\ell) +\frac{3\sqrt{N}\beta(k_0)}{2^{b+1}} \nonumber \\
    &\le \frac{\beta(k_0+1)}{2}.
\end{align}
Given that $\alpha_i(k_0+1)=\xi_i^q(k_0-1) - \beta(k_0+1)/2$ in \eqref{eq:alpha_dynamics}, we obtain from \eqref{eq:xi_final} that $\xi_i(k_0) \in [\alpha_i(k_0+1), \alpha_i(k_0+1)+\beta(k_0+1) ]$. 
Thus the lemma holds for $k_0+1$ and the proof is complete. \qed

\section{Proof of Theorem \ref{T:main}} \label{Ap:main}
Let us define the quantity $\phi(k)=x(k+1)-Lx(k)$. 
From \eqref{eq:xi_quant_2} we have
\begin{align}
    x(k+1) = Lx(k) - \Delta^q(\xi(k),k+1), \label{eq:x_dynamics_2}
\end{align}
and hence, $\phi(k)= - \Delta^q(\xi(k),k+1).$
Therefore,
\begin{align*}
    \lim_{k\to \infty}\|\phi(k)\|=\|\Delta^q(\xi(k),k+1)\|\le \lim_{k\to \infty} \frac{\beta(k+1)}{2^{b+1}},
\end{align*}
where we used \eqref{eq:delta_bound}.
Since $b$ is chosen such that \eqref{eq:b_eta} is satisfied, we have that $\lim_{k\to \infty} \beta(k)=0$ due to Lemma~\ref{L:beta_rate}. 
Therefore, $\phi(k)$ converges to $0$ and $x(k)$ converges to a point $\tilde{x}_\infty$ such that $\tilde x_\infty=L \tilde x_\infty$. 
Since the right eigenvector of $L$ corresponding to the eigenvalue $1$ must be of the form $c\textbf{1}$ for some constant $c\in \R$, we have $\tilde x_\infty=c\textbf{1}$ for some $c$, and hence, consensus is achieved.

Recall that without quantization the consensus is achieved at the point $x_\infty=\textbf{1}(w_\ell\T {x}_0)$, where recall that $w_\ell$ is the left eigenvector of $L$ and ${x}_0=[x_{10},\ldots,x_{N0}]$ is the initial state of the agents.
Although, $\tilde{x}_\infty$ may differ from $x_\infty$, we can bound the difference between $x_\infty$ and $\tilde x_\infty$ as follows.
Use \eqref{eq:x_dynamics_2} to write
\begin{align*}
    x(k) =& L^k x(0) - \sum_{t=0}^{k-1}L^{k-1-t}\Delta^q(\xi(t),t+1)\\
    =&L^k ({x}_0 + \Delta^q({x}_0,0)) - \sum_{t=0}^{k-1}L^{k-1-t}\Delta^q(\xi(t),t+1).
\end{align*}
Now, recall that $x_\infty = \lim_{k\to \infty} L^k {x}_0$ and hence, from the last equation,
\begin{align*}
    \|\tilde x_\infty -x_\infty\|&=\|\lim_{k\to \infty}x(k)-\lim_{k\to \infty }L^k {x}_0\|\nonumber \\
   & =\lim_{k\to \infty }\|x(k)-L^k {x}_0\|\\
   & \overset{(a)}{\le} \|\Delta^q({x}_0,0)\| +\lim_{k \to \infty}\sum_{t=0}^{k-1}\|\Delta^q(\xi(t),t+1)\| \\
    &\overset{(b)}{\le} \frac{\sqrt{N}\beta(0)}{2^{b+1}}+\lim_{k\to \infty }\frac{\sqrt{N}\bar\beta}{2^{b+1}}\frac{~\eta-\eta^{k+1}}{\!\!1-\eta}\\
    &=\left(\beta(0)+\frac{\bar\beta\eta}{1-\eta}\right) \frac{\sqrt{N}}{2^{b+1}},
\end{align*}
where to obtain inequality $(a)$, we have used the fact that $\|L\|\le 1$, and in deriving inequality $(b)$, we have used \eqref{eq:delta_bound} and Lemma~\ref{L:beta_rate}.
This completes the proof.\qed

\section{Proof of Theorem \ref{T:finall}} \label{AP:T_finall}

To obey the finite bit-rate constraint, agent $i$ will send a quantized version of $\xi_i(k)$ to its outgoing neighbors.
Therefore, the adversary $i$ intercepting the inter-agent communications will have access to $\xi_i^q(k)$ instead of $\xi_i(k)$.
Following similar steps as in the proof of Theorem \ref{T:ICC}, one may verify that the adversary $i$ will follow the estimation dynamics (see also \eqref{eq:ap_adv_est_dyn2}) 
\begin{align*}
    \xad_i(k+1) =  \xad_i(k)+\mu_i(k)\xi_i^q(k).
\end{align*}
Consequently, the estimation error $\ead(k)=x(k)-\xad(k)$ will have the dynamics
\begin{align*}
    \ead(k+1)=\ead(k)+M(k)\xi^q(k), \quad \ead(-1) = 0
\end{align*}
where recall that $M(k)= I- \diag{\mu_1(k),\cdots,\mu_N(k)}$.
The mean $\E[\ead(k)]$ and the second moment $\sad(k)=\E[\ead(k)\ead(k)\T]$ of the estimation error follow the dynamics
\begin{subequations} \label{eq:quantization_dynamics_adversary}
\begin{align}
    &\E[\ead(k+1)]=\E[\ead(k)]+(1-\gamma) \xi^q(k), \label{eq:quantization_ead}\\
    & \E[\ead(-1)] = 0 \nonumber \\
   & \sad(k+1)=\sad(k) + \Gamma(k), \quad \sad(-1) = 0,\label{eq:quantization_sigma}\\
   & \Gamma(k)=(1-\gamma)\Big(\E[\ead(k)] \xi^q(k)\T \nonumber +\xi^q(k)\E[\ead(k)]\T\Big)\\
   &~~~~~\quad +(1-\gamma )^2\xi^q(k)\xi^q(k)\T\nonumber \\
   &~~~~~\quad+\gamma(1-\gamma) \diag{\xi_1^q(k)^2,\ldots,\xi_N^q(k)^2}, \label{eq:quantization_gamma} 
\end{align}
\end{subequations}
where in deriving the expression of $\Gamma(k)$ we have used the fact that   $\E[\mu_i(k)^2]=\gamma$ and $\E[\mu_i(k)\mu_j(k)]=\gamma^2$ for $i\ne j$ and for all $k\in\bbN_0$.
From \eqref{eq:xi_quant_2} notice that $x(k+1) =x(k) + \xi^q(k)$ and thus, $x(k)=\sum_{t=-1}^{k-1}\xi^q(k)$ for all $k$.
Therefore, from \eqref{eq:quantization_ead} we may write,
\begin{align}
    \E[\ead(k+1)]=(1-\gamma)\sum_{t=-1}^k\xi^q(t)=(1-\gamma)x(k+1). \label{eq:ead_x}
\end{align}
Using \eqref{eq:ead_x}, we simplify \eqref{eq:quantization_gamma} as follows
\begin{align*}
    \Gamma(k)= &(1-\gamma)^2(x(k)\xi^q(k)+\xi^q(k)x(k)\T+\xi^q(k)\xi^q(k))\\
    &+\gamma(1-\gamma) \diag{\xi_1^q(k)^2,\ldots,\xi_N^q(k)^2}\\
    =&(1-\gamma)^2(x(k+1)x(k+1)\T-x(k)x(k)\T) \\
    &+\gamma(1-\gamma) \diag{\xi_1^q(k)^2,\ldots,\xi_N^q(k)^2},
\end{align*}
for all $k\in \bbN_0$, and  for $k=-1$, \eqref{eq:quantization_gamma} yields
\begin{align*}
    \Gamma(-1)= &(1-\gamma)^2\xi^q(-1)\xi^q(-1)\T\\
    &+\gamma(1-\gamma) \diag{\xi_1^q(-1)^2,\ldots,\xi_N^q(-1)^2}\\
    =&(1-\gamma)^2x(0)x(0)\T \\
    &+\gamma(1-\gamma) \diag{(x_{10}^q)^2,\ldots,(x_{N0}^q)^2}.
\end{align*}
Using this simplified expression for $\Gamma(k)$, we obtain from \eqref{eq:quantization_sigma} that
\begin{align*}
    \sad(k+1)=&\sum_{t=-1}^{k}\Gamma(t) \\
    =&(1-\gamma)^2x(k+1)x(k+1)\T \\
    &+ \gamma(1-\gamma)\sum_{t=-1}^k  \diag{\xi_1^q(k)^2,\ldots,\xi_N^q(k)^2} \\
    \succeq &~ (1-\gamma)^2x(k+1)x(k+1)\T.
\end{align*}
Therefore, $ \E[\ead(k)\T \ead(k)] \ge (1-\gamma)^2\|{x}(k)\|^2+c$, where $c=\gamma(1-\gamma)\|x_{i0}^q\|^2$, which proves that the ICC algorithm is \ep{(1-\gamma)^2} even under the bit-rate constraint. 
From Definition \eqref{Df:e-protected} $\epsilon$-protection is a sufficient condition for asymptotic $\epsilon$-protection, and hence the ICC algorithm is \aep{(1-\gamma)^2} under bit rate constraint as well. \qed


\bibliographystyle{IEEEtran}
\bibliography{references}

\end{document}